\documentclass[aps,prb,twocolumn,floats,epsfig]{revtex4}
\usepackage[iso-8859-1]{inputenx}
\usepackage{amssymb}
\usepackage{amsbsy}
\usepackage{amsmath}
\usepackage{epsfig}
\usepackage{color}
\usepackage{float}
\usepackage[colorlinks,linktocpage,bookmarks=false,citecolor=blue,linkcolor=red,urlcolor=blue]{hyperref}

\newcommand{\ing}{\includegraphics}
\newcommand{\bib}{\bibitem}
\newcommand{\beq}{\begin{equation}}
\newcommand{\eeq}{\end{equation}}
\newcommand{\bea}{\begin{eqnarray}}
\newcommand{\eea}{\end{eqnarray}}

\usepackage{xcolor}

\begin{document}

\title{Periodically driven Rydberg chains with staggered detuning}

\author{Bhaskar Mukherjee$^1$, Arnab Sen$^2$, K. Sengupta$^2$}

\affiliation{$^1$ Department of Physics, University College London,
Gower Street, London WC1E 6BT, UK. \\
$^2$School of Physical Sciences, Indian Association for the
Cultivation of Science, Jadavpur, Kolkata 700032, India}

\date{\today}

\begin{abstract}

We study the stroboscopic dynamics of a periodically driven finite
Rydberg chain with staggered ($\Delta$) and time-dependent uniform
($\lambda(t)$) detuning terms using exact diagonalization (ED). We
show that at intermediate drive frequencies ($\omega_D$), the
presence of a finite $\Delta$ results in violation of the eigenstate
thermalization hypothesis (ETH) via clustering of Floquet
eigenstates. Such clustering is lost at special commensurate drive
frequencies for which $\hbar \omega_d=n \Delta$ ($n \in Z$) leading
to restoration of ergodicity. The violation of ETH in these driven
finite-sized chains is also evident from the dynamical freezing
displayed by the density-density correlation function at specific
$\omega_D$. Such a correlator exhibits stable oscillations with
perfect revivals when driven close to the freezing frequencies for
initial all spin-down ($|0\rangle$) or Neel ($|{\mathbb
Z}_2\rangle$, with up-spins on even sites) states. The amplitudes of
these oscillations vanish at the freezing frequencies and reduces
upon increasing $\Delta$; their frequencies, however, remains pinned
to $\Delta/\hbar$ in the large $\Delta$ limit. In contrast, for the
$|{\bar {\mathbb Z}_2}\rangle$ (time-reversed partner of $|{\mathbb
Z}_2\rangle$) initial state, we find complete absence of such
oscillations leading to freezing for a range of $\omega_D$; this
range increases with $\Delta$. We also study the properties of
quantum many-body scars in the Floquet spectrum of the model as a
function of $\Delta$ and show the existence of novel mid-spectrum
scars at large $\Delta$. We supplement our numerical results with
those from an analytic Floquet Hamiltonian computed using Floquet
perturbation theory (FPT) and also provide a semi-analytic
computation of the quantum scar states within a forward scattering
approximation (FSA). This allows us to provide qualitative
analytical explanations of the above-mentioned numerical results.

\end{abstract}

\maketitle

\section{Introduction}

The physics of non-equilibrium quantum systems has been studied
extensively in recent years
\cite{rev1,rev2,rev3,rev4,rev5,rev6,rev7,rev8}. The theoretical
studies in the field have been boosted by possibility of
experimental realization using ultracold atom platforms
\cite{rev9,exp1,exp2,exp3,exp4}. Out of the several possible drive
protocols that could be used to take a quantum system out of
equilibrium, periodic drives have attracted a lot of recent
attention \cite{rev6,rev7,rev8}. For such protocols characterized by
a time period $T$, the unitary evolution operator $U(t,0)$ governs
the dynamics of the system. At times $t= nT$, (where $n\in Z$) $U$
can be written in terms of the Floquet Hamiltonian $H_F$: $U(nT,0)=
\exp[-i H_F n T/\hbar]$, where $\hbar$ is Planck's constant. Thus
the stroboscopic dynamics of the system is completely governed by
its Floquet Hamiltonian \cite{rev8}.

The interest in such drive protocols stems from several interesting
physical phenomena which occurs in periodically driven systems.
These include drive-induced non-trivial topology
\cite{topo1,topo2,topo3}, dynamical localization
\cite{dloc1,dloc2,dloc3,dloc4}, dynamical phase transitions
\cite{dtran1,dtran2,dtran3}, realization of time crystals
\cite{tc1,tc2,tc3}, dynamical freezing
\cite{df1,df2,df3,df4,df5,ethv2} and the possibility of tuning
ergodicity property of the driven system \cite{ethv1}. More
recently, some of these properties have also been investigated in
the context of quasiperiodically \cite{qpd1,qpd2,rand1} and
aperiodically \cite{rand1} driven systems.

A central paradigm for understanding the behavior of non-integrable
many-body systems described by local Hamiltonians that are driven
out of equilibrium comes from the eigenstate thermalization
hypothesis (ETH) \cite{eth1,eth2,eth3,eth4}. ETH postulates that the
reduced density matrix of any eigenstate with finite energy density
is thermal in the thermodynamic limit, the corresponding temperature
being determined by its energy density. ETH has also proved
successful in explaining properties of non-integrable systems in the
presence of a periodic drive; here eigenstates of the Floquet
Hamiltonian play a similar role \cite{eth4}.

The violation of ETH in many-body systems can occur via several
routes \cite{mbl1,mblrev1,mblrev2,mblrev3,mblrev4,
ethv0,ethv01,ethv02,ethv03,ethv04,ethv05a,ethv05b,
ethv06,ethv07,ethv1,ethv2,ethv08,ethv09,hsf1,hsf2,hsf3,hsf4}. For
example, ETH does not hold for integrable models where the presence
of an extensive number of additional conserved quantities makes the
system non-ergodic. Moreover, a loss of ergodicity and hence
violation of ETH may also happen due to strong disorder when a
system enters a many-body localized (MBL) phase
\cite{mbl1,mblrev1,mblrev2,mblrev3,mblrev4}. Such violation may also
occur due to Hilbert space fragmentation where the system Hilbert
space organizes into dynamically disconnected sectors
\cite{hsf1,hsf2,hsf3,hsf4}.

Another weaker version of ETH violation has also been found in
Rydberg atom systems \cite{ethv0, ethv01,ethv02,ethv03,ethv05b,
ethv1,ethv2}. Experiments on such systems showed coherent long-lived
oscillation of density of Rydberg excitations during evolution
starting from a Neel state with one Rydberg excitation at every even
($|{\mathbb Z}_2 \rangle = |..1_{2j},0_{2j+1},1_{2j+2},....\rangle$)
or odd ($|{\bar {\mathbb Z}_2}\rangle = |... 1_{2j-1}, 0_{2j},
1_{2j+1}, ....\rangle$) site; in contrast, ETH predicted
thermalization occurred when the dynamics started from the Rydberg
vacuum state ($|0\rangle= |..0_j,0_{j+1},\rangle$) \cite{exp4}. The
reason for such ETH violating oscillations were traced to the
existence of a special class of eigenstates called quantum scars
\cite{ethv0,ethv01,ethv02,ethv03,ethv04,ethv05a,ethv05b,
ethv06,ethv07,ethv1,ethv2,ethv08,ethv09}. These are many-body
eigenstates with finite energy density but anomalously low
entanglement entropy. Consequently, dynamics starting from an
initial state which has large overlap with them do not show
thermalization. The role of such states in periodically driven
Rydberg chain has also been analyzed \cite{ethv1,ethv2}; it was
shown that their presence may lead to the possibility of
drive-frequency induced tuning ergodicity property of these chains
\cite{ethv1}, presence of subthermal steady states, and the
phenomenon of dynamical freezing \cite{ethv2}. These properties do
not have any analogue for scar-induced dynamics following a quench.

More recently, there have been several studies on possible role of
confinement in condensed matter systems. The concept of confinement
in 1D quantum electrodynamics is well-known; it was shown that such
systems are characterized by a parameter $\theta$ which is related
to the background electric field. For $\theta=\pi$, the system is
deconfined and charge excitations can propagate; for other values of
$\theta$, these excitations are confined \cite{qedcon1,qedcon2}.
Such ideas have also been applied to the simulation of lattice gauge
theories (LGTs) using ultracold atom platforms
\cite{simu1,simu2,simu3,simu4,deb1,deb2}, Ising like chains
\cite{conf1}, and Rydberg chains with staggered detuning described
by the Hamiltonian \cite{ssg,fss,ethv0,conf2,conf3}
\begin{eqnarray}
H_{\rm ryd} &=& -w \sum_j \tilde \sigma_j^x - \frac{1}{2} \sum_j
(\lambda + \Delta (-1)^j) \sigma_j^z  \label{rydham1}
\end{eqnarray}
Here $\Delta>0$ and $\lambda$ denote the amplitude of staggered and
uniform detuning respectively, $w>0$ is the coupling strength
between ground and excited states of the Rydberg atoms, $\sigma_j^z=
2\hat n_j-1$ is related to the number operator $\hat n_j$ for a
Rydberg excitations on site $j$. The Hamiltonian given by Eq.\
\ref{rydham1} is to be supplemented by a constraint that two
adjacent sites can not both have Rydberg excitations. Thus the model
has a Rydberg blockade radius of one lattice site. This constraint
is implemented using the projected spin operators $\tilde \sigma_j^x
= P_{j-1} \sigma_j^x P_{j+1}$, where $P_{j-1} = (1-\sigma_j^z)/2$.
The dynamics of such Rydberg chains was studied in context of quench
protocols in Ref.\ \onlinecite{conf2,conf3}. It was observed that
the presence of the confinement terms leads to slow dynamics of the
Rydberg atoms. However, periodic drive protocols for such chains
have not been studied so far.

In this work, we study a periodically driven Rydberg chain with a
staggered detuning term. Here we consider the uniform detuning
parameter $\lambda$ to be a periodic function of time given by
\begin{eqnarray}
\lambda(t) &=& \lambda_0 \quad  t\le T/2 \nonumber\\
&=& -\lambda_0  \quad t > T/2 \label{proto1}
\end{eqnarray}
where $T=2\pi/\omega_D$, $\omega_D$ is the drive frequency and we
focus in the regime of large drive amplitude:$\lambda_0 \gg \Delta,
w$. In the rest of this work we focus on the dynamics of the
density-density correlation function ${\rm O}_{j 2} = \langle \hat
n_j \hat n_{j+2} \rangle$ of these driven Rydberg atoms and the
half-chain entanglement entropy $S_{L/2}$ of their Floquet
eigenstates. We use ED to study the properties of $S_{L/2}$ and
${\rm O}_{j2}$ numerically. We supplement our numerical finding by
comparing them with those obtained from an analytic, albeit
perturbative, Floquet Hamiltonian derived using Floquet perturbation
theory (FPT) and also present a calculation of some Floquet
eigenstates using forward scattering approximation (FSA) applied to
the analytic Floquet Hamiltonian.

The main results that we obtain from such a study are as follows.
First, we show, via study of $S_{L/2}$, the presence of clustering
of Floquet eigenstates in these systems for strong staggered
detuning and at intermediate frequencies leading to violation of
ETH. These clusters can be classified into two categories. The first
type, dubbed as primary clusters, is a result of near-integrable
nature of the system for large staggered detuning. The second type
of clusters occurs within each primary cluster at a much smaller
quasienergy scale. We show that these secondary clusters result from
an emergent conservation law found earlier in non-driven staggered
Rydberg chains \cite{ethv05b}. Second, we also show that the
secondary clustering is destroyed, leading to ergodicity restoration
within each primary cluster, at commensurate drive frequencies
$\hbar \omega_D =n \Delta$, where $n$ is an integer; this allows
drive-induced tuning of ergodicity property of the Floquet
eigenstates within each primary cluster. Third, we show that for
almost all intermediate drive frequencies, ${\rm O}_{22}$ does not
reach its ETH predicted steady state value; this phenomenon is seen
for $|{\mathbb Z}_2\rangle$, $|0\rangle$ and the $|{\bar {\mathbb
Z}_2}\rangle$ initial states. For dynamics starting from $|{\mathbb
Z}_2\rangle$ and $|0\rangle$ initial states, we find specific drive
frequencies for which the system is dynamically frozen; ${\rm
O}_{22}$ remains pinned to its initial values at these frequencies.
Around these freezing frequencies, ${\rm O}_{22}$ displays
oscillatory behavior around its initial values with perfect
revivals; the amplitudes of these oscillations vanish as the
freezing frequency is approached and reduces as $\Delta$ is
increased. In contrast, the frequency of these oscillation are
pinned to $\Delta/\hbar$ for a wide range of $\Delta$ in the
intermediate and large $\Delta$ regime. We provide a semi-analytic
explanation of this phenomenon. Fourth, in contrast to the above
cases, for the $|{\bar {\mathbb Z}_2}\rangle$ initial state,
$O_{22}$ remains frozen over a range of frequencies leading to a
dynamically frozen phase. We provide a qualitative explanation for
this phenomenon; our analysis shows that for odd $j$, $O_{j2}$ will
have a frozen phase for dynamics starting from $|{\mathbb
Z}_2\rangle$. This dichotomy is a direct consequence of the $Z_2$
(between even and odd sites) symmetry breaking due to the presence
of the staggered detuning. Fifth, we study the fate of the quantum
many-body scars in the Floquet eigenspectrum as a function of the
staggered detuning. For small $\Delta$, these scars have almost
equal overlap with both the Neel states; in contrast, we find that
scars with $E_F<0$ ($E_F>0$) have stronger overlap with $|{\mathbb
Z}_2\rangle$ ($|{\bar {\mathbb Z}_2}\rangle$). For large enough
$\Delta$, the mid spectrum scars ($E_F \simeq 0$) do not show
significant overlap with either of the Neel states; instead we show
that these represent a separate type of scar states exhibiting high
overlap either with number states with broken $Z_4$ symmetry
($|{\mathbb Z}_4\rangle$) or with states having single Rydberg
excitation ($|1\rangle$). These scar states results in long-time
coherent oscillations for dynamics starting from $|{\mathbb
Z}_4\rangle$ or $|1\rangle$ initial states and are therefore
qualitatively different from their counterparts in the PXP model
\cite{ethv01}. We also present a computation of the scars at small
and intermediate $\Delta$ using a FSA formalism \cite{turpap1}.

The plan of the rest of the paper is as follows. In Sec.\ \ref{fpt},
we obtain the analytic Floquet Hamiltonian using FPT. This is
followed by Sec.\ \ref{sd} where we study the structure of the
Floquet eigenspectrum for strong staggered detuning and discuss the
resultant clustering. Next, in Sec.\ \ref{df}, we analyze the
dynamical freezing of $O_{22}$ and its dynamics near the freezing
points. This is followed by Sec.\ \ref{crover} where we discuss the
change in properties of the quantum many-body scars in the Floquet
eigenspectra as $\Delta$ is tuned from small to large values.
Finally, we summarize our main results and conclude in Sec.\
\ref{diss}. Some details of the calculation are presented in the
appendix.

\section{Derivation of the Floquet Hamiltonian}
\label{fpt}

In this section, we derive an analytic Floquet Hamiltonian for the
driven Rydberg atoms described by Eq.\ \ref{rydham1} in the presence
of the drive protocol given by Eq.\ \ref{proto1}. To this end, we
shall focus on the limit $\lambda_0 \gg \Delta, w$; however, no
restriction is placed on the relative magnitude of $\Delta$ and $w$.
For the rest of this section, we shall set $\hbar=1$.

In what follows, we shall use FPT to obtain the Floquet Hamiltonian
\cite{rev9,ethv2}. Since $\lambda_0 \gg \Delta, w$, we treat the
uniform detuning term exactly and rewrite Eq.\ \ref{rydham1} as $H
=H_0(t) + H_1+H_2$ where
\begin{eqnarray}
H_0(t) &=& -\frac{1}{2} \lambda(t) \sum_j \sigma_j^z, \nonumber\\
H_{1} &=& - \frac{1}{2} \sum_j \Delta (-1)^j \sigma_j^z, \quad
H_{2}= - w \sum_j \tilde \sigma_j^x. \label{pertham}
\end{eqnarray}
We shall treat $H_1$ and $H_2$ perturbatively. This leads to the
evolution operator
\begin{eqnarray}
U_0(t,0) &=&  T_t \left[e^{-i \int_0^t H_0(t') dt'}\right] \nonumber\\
&=& e^{i \lambda_0 t \sum_j \sigma_j^z /2}  \quad t \le T/2
\nonumber\\
&=& e^{i \lambda_0 (T-t) \sum_{j} \sigma_j^z /2} \quad t \ge T/2,
\label{zerofl}
\end{eqnarray}
We note that $U_0(T,0)=I$ which leads to $H_F^{(0)}=0$. Also, for
subsequent computation, we define the states $|m\rangle$ as
eigenstates of $H_0$ that have $m$ up-spins and $L-m$ down-spins.
Such states need to obey the constraint imposed by the model; thus
$m_{\rm max}=L/2$ is the maximum number of up-spins.

Next, we consider the first order correction to the Floquet
Hamiltonian given, within FPT, by
\begin{eqnarray}
U_1(T,0) &=& -i \int_0^T dt_1 U_0^{\dagger}(t_1,0) (H_1+H_2)
U_0(t_1,0) \label{forderU}
\end{eqnarray}
To evaluate $U_1$, we note that $H_{1}$ commutes with $U_0$.
Moreover, the contribution from $H_{2}$ has already been obtained in
Refs.\ \onlinecite{ethv1,ethv2}. From these we find,
\begin{widetext}
\begin{eqnarray}
\langle m|U_1(T,0)|n\rangle  &=&  i T \sum_m \left( \sum_j
\frac{\Delta}{2} (-1)^j \langle m|\sigma_j^z|m\rangle |m\rangle
\langle m| \delta_{mn} + \sum_{s_j=\pm 1} \frac{4 i w}{\lambda_0 T}
\sin \frac{\lambda_0 T}{4} e^{i \lambda_0 T s_j} |m\rangle \langle
m+s_j| \delta_{n, m+s_j} \right)  \label{ufirst}
\end{eqnarray}
\end{widetext}
Using Eq.\ \ref{ufirst}, one obtains the first order Floquet
Hamiltonian, given by $H_F^{(1)}= i U_1(T,0)/T$  to be
\begin{eqnarray}
H_F^{(1)} = H_{1} - w_{r} \sum_j \left(\cos \gamma \tilde \sigma_x +
\sin \gamma \tilde \sigma_y \right), \,\, w_r= \frac{w
\sin\gamma}{\gamma}
\end{eqnarray}
where $\gamma=\lambda_0 T/4$. We note that the second term in
$H_{F}^{(1)}$, obtained in Refs.\ \onlinecite{ethv1,ethv2}, vanishes
at $\gamma= n \pi$ (or $\lambda_0 = 2 n \omega_D$) for non-zero
integer $n$; thus at these points, we expect the effect of the
staggered detuning term ($H_1$) to be particularly strong.

It turns out that there is no second order contribution to $H_F$
since
\begin{eqnarray}
U_2(T,0)&=& - \int_0^T dt_1 U_0^{\dagger}(t_1,0) (H_1 +H_2)
U_0(t_1,0)
\nonumber\\
&& \times \int_0^{t_1} dt_2 U_0^{\dagger}(t_2,0) (H_1+H_2)
U_)(t_2,0) \nonumber\\
&=& U_1(T,0)^2/2 \label{secondord}
\end{eqnarray}
Eq.\ \ref{secondord} can be obtained from a straightforward
computation; however, it can also be checked as follows. First we
note that the terms $\sim H_1$ in the expression of $U_2$ commutes
with $U_0$. This indicates that the term originating from $H_1^2$ in
$U_2(T,0)$ must trivially satisfy Eqn.\ \ref{secondord}. Second, we
recall that it is known \cite{ethv1,ethv2} that the term $\sim
H_2^2$ does not yield any second order contribution to $H_F$. This
is due to the presence of a symmetry noted in Ref.\
\onlinecite{ethv1}. For $\Delta=0$, it was shown that $\{Q,
H_F\}=0$, where $Q= \prod_j \sigma_j^z$; consequently, there can be
no terms with product of even number of $\tilde \sigma_j^{\pm}$
operators in $H_F$. Moreover, a straightforward computation shows
that the contribution from all terms which involve both $H_1$ and
$H_2$ also vanish due to commutation of $H_1$ with $U_0$. Thus we
have $H_F^{(2)}=0$.

Finally, we compute the third order term in the Floquet Hamiltonian.
To this end, we consider the third order contribution to $U$, given
by
\begin{eqnarray}
U_3(T,0) &=& i \int_0^{T} dt_2 U_0^{\dagger}(t_1,0) (H_1+H_2)
U_0(t_1,0)
\nonumber\\
&& \times \int_0^{t_1} dt_2 U_0^{\dagger}(t_2,0) (H_1+H_2)
U_0(t_2,0)\nonumber\\
&& \times \int_0^{t_2} dt_2 U_0^{\dagger}(t_3,0) (H_1+H_2)
U_0(t_3,0).  \label{thirdorder}
\end{eqnarray}
To obtain the third order Floquet Hamiltonian, we first note that
the commutation of $H_1$ with $U_0$ ensures that there is no third
order contribution to $H_F$ from the term with three $H_1$
operators. The first non-trivial contribution to the third order
$H_F$ comes from the term in Eq.\ \ref{thirdorder} which has two
$H_1$ and one $H_2$ operators. The contribution of this term to
$U_{3}(T,0)$ can be computed as charted in details in Ref.\
\onlinecite{ethv2}. The result is
\begin{eqnarray}
U_{3a}(T,0)&=& \frac{i \Delta_0^2 w}{4} \sum_{j_1,j_2,j_3}
\sum_{s=\pm} (-1)^{j_1+j_2} ( 2 c_{1s} \sigma_{j_1}^z \sigma_{j_2}^z
\tilde \sigma_{j_3}^s \nonumber\\
&& + c_{2s}
\sigma_{j_1}^z \tilde \sigma_{j_3}^s \sigma_{j_2}^z) \label{u3aexp} \\
c_{1\pm} &=& \pm T^3 \frac{-i \pm 2 \gamma  (1 \pm 2 i\gamma)-i
e^{\pm 2i \gamma}\left(2 \gamma^2-1\right)}{32 \gamma^3}
\nonumber\\
c_{2 \pm} &=& \pm T^3 \frac{i \mp 2 \gamma  -i e^{\pm 2 i\gamma
}\left(2\gamma^2 +1\right)}{16 \gamma^3} \nonumber
\end{eqnarray}
The structure of $U_{3a}$ shows that it is necessary to categorize
the terms into two groups. The first constitutes the case where at
least one of the $j_i$s are different from the other two. For these,
$U_{3a}(T,0)$ either vanishes or can be written as
\begin{eqnarray}
U_{3a}(T,0) &=& \frac{i \Delta_0^2 w}{4} \sum_{j_1 \ne j_2 \ne j_3}
\sum_{s=\pm} (-1)^{j_1+j_2} (2 c_{1s} +c_{2s})
\nonumber\\
&& \times \sigma_{j_1}^z \sigma_{j_2}^z \tilde \sigma_{j_3}^s =
\frac{U_1(T,0)^3}{3!}
\end{eqnarray}
where we have used the fact that $2c_{1s}+c_{2s}= i
T^2(\exp[i\lambda_0 T s]-1)/(\lambda_0 s)$ for $s=\pm 1$. Thus such
terms do not contribute to $H_{F3}$. However, when $j_1=j_2=j_3$, we
find that the last term in Eqn.\ \ref{u3aexp} acquires a negative
sign due to the anticommutation of $\sigma_j^z$ and $\tilde
\sigma_j^s$. This leads to
\begin{eqnarray}
U_{3a}(T,0) &=& \frac{i \Delta_0^2 w}{4} \sum_{j} \sum_{s=\pm} (2
c_{1s} - c_{2s}) \tilde \sigma_j^s
\end{eqnarray}
Comparing this with $U_1(T,0)^3/3!$, we find that there is a
non-trivial contribution to the Floquet Hamiltonian given by
\begin{eqnarray}
H_{F1}^{(3)} &=& \frac{w \Delta_0^2}{2T}  \sum_j (c_{2+} \tilde
\sigma_j^+ +{\rm h.c.}) \label{tordfl1}
\end{eqnarray}
This term leads to a frequency dependent renormalization of the
coefficient of $w$ in the first order Floquet Hamiltonian.

The next contribution to the Floquet Hamiltonian comes from the
terms in Eqn.\ \ref{thirdorder} with two $H_2$ and one $H_1$
operators. The calculation of this term, $U_{3b}(T,0)$, follows a
similar route as that followed in Ref.\ \onlinecite{ethv2} and leads
to
\begin{widetext}
\begin{eqnarray}
U_{3b}(T,0) &=& \frac{i \Delta_0 w^2}{2} \sum_{j_1,j_2,j_3}
\sum_{s_1,s_2=\pm} (-1)^{j_3} ( 2 d_{1 s_1 s_2} \tilde
\sigma_{j_1}^{s_1} \tilde \sigma_{j_2}^{s_2} \sigma_{j_3}^z + d_{2
s_1 s_2}
\tilde \sigma_{j_1}^{s_1} \sigma_{j_3}^z \tilde \sigma_{j_2}^{s_2}) \label{u3bexp} \\
d_{1 + +} &=& T^3 \frac{-(i/4+ \gamma ) + i e^{2 i\gamma}(i + 2
\gamma)+ e^{12 i \gamma} (3 i/4 + 2 \gamma)}{32 \gamma^3},
\nonumber\\
d_{1 + -} &=& T^3 \frac{\gamma \left(2- i \gamma\right) - \gamma
e^{2 i
\gamma} - i\left(1-e^{2i \gamma}\right)/2}{16 \gamma^3} \nonumber\\
d_{2 + +} &=& T^3 \frac{i/4 - \gamma + 3 e^{4 i\gamma}/4 + e^{2i
\gamma} (-i + 2 \gamma)}{16 \gamma^3}, \quad d_{2 + -} =
\frac{T^3}{8 \gamma^3} \left(\frac{1}{2} \sin 2\gamma - \gamma \cos
2\gamma\right)
\end{eqnarray}
\end{widetext}
and $d_{a \pm \pm}=d_{a \mp \mp}^{\ast}$ for $a=1,2$. The argument
leading to the Floquet Hamiltonian is similar to the one discussed
earlier and only the on-site terms in $U_{3b}$ contribute to
$H_F^{(3)}$. The final result is
\begin{eqnarray}
H_{F2}^{(3)} &=& -\frac{w^2\Delta_0}{ T} \sum_j (-1)^j (d_{2+-}
\tilde \sigma_j^+ \sigma_j^z \tilde \sigma_j^- +{\rm h.c.} )
\label{tordfl2}
\end{eqnarray}
Note that the action of $H_{F2}^{(3)}$ on any eigenstate of
$\sigma_j^z$ is same as that of $-\sigma_j^z$. Thus this term can be
considered as a correction of the staggered detuning term in the
first order Floquet Hamiltonian. From Eq.\ \ref{u3bexp}, we find
that for $\tan 2\gamma
> 2\gamma $, this correction is negative and leads to a
reduction in the magnitude of the on-site detuning.

The final term in the Floquet Hamiltonian comes from the term in
Eq.\ \ref{thirdorder} which has three $H_2$ operators. This term has
already been computed in Ref.\ \onlinecite{ethv2}; for completeness,
we write the contribution of this term to $H_F$ here. This is given
by
\begin{eqnarray}
H_{F3}^{(3)} &=& \sum_j A_0 [(\tilde \sigma_{j-1}^+ \tilde
\sigma_{j+1}^+ + \tilde \sigma_{j+1}^+ \tilde \sigma_{j-1}^+ )\tilde
\sigma_{j}^- - 6 \sigma_j^+] + {\rm h.c} \nonumber\\
A_0 &=&  [e^{6 i\gamma} + 3 e^{2 i \gamma} (1 + 4 i\gamma) + 2(1 - 3 e^{12 i \gamma} )] \nonumber\\
&& \times \frac{w^3 T^2 e^{-12 i \gamma}}{192 i \gamma^3}
\label{tordfl3}
\end{eqnarray}
This term contains a three-spin term as well as a higher-order
correction to the first order Floquet term $\sim w$.

This completes our derivation of the Floquet Hamiltonian for the
driven Rydberg chain with staggered detuning to third order in
Floquet perturbation theory. In subsequent sections, we shall use
this along with numerical results to analyze the properties of the
driven chain.

\section{Strong staggered detuning}
\label{sd}

In this and the next two sections, we present the main results of
our study obtained by exact numerics and compare them to that
obtained from the analytic Floquet Hamiltonian derived in Sec.\
\ref{fpt}. In the present section, we shall restrict ourselves to
the strong staggered detuning limit ($\Delta\ge w$), while other
parameter regimes will be explored in Secs.\ \ref{df} and
\ref{crover}.

The exact numerical results that we present are obtained using ED on
finite sized chain as follows. We first numerically diagonalize
$H[\pm \lambda_0]=H_0[\pm \lambda_0]+H_1+H_2$ (Eq.\ \ref{pertham})
to obtain their energy eigenvalues $\epsilon_a^{\pm}$ and
eigenvectors $|\xi_a^{\pm}\rangle$. The evolution operator for the
square pulse protocol (Eq.\ \ref{proto1}) can then be written as
\begin{eqnarray}
U(T,0) &=& e^{-i H_F[-\lambda_0] T/(2\hbar)} e^{-i H_F[\lambda_0]
T/(2\hbar)} \nonumber\\
&=& \sum_{a,b} c_{a b}^{-+} e^{-i(\epsilon_a^-+\epsilon_b^+)
T/(2\hbar)} |\xi_a^-\rangle \langle \xi_b^+| \label{unum}
\end{eqnarray}
where $c_{ab}^{-+} = \langle \xi^-_a| \xi_b^+\rangle$. We then
numerically diagonalize the unitary matrix $U$ to obtain its
eigenvalues $\lambda_n = \exp[-i E_n^F T/\hbar]$ and eigenvectors
$|\chi_n\rangle$. The exact Floquet Hamiltonian of the system can
then be obtained as
\begin{eqnarray}
H_F &=& \sum_n E_n^F |\chi_n\rangle \langle \chi_n | \label{flnum}
\end{eqnarray}
In what follows, we shall use Eq.\ \ref{flnum} for numerical
computation of the Floquet Hamiltonian. Most the numerical results
in the rest of this section, unless explicitly mentioned otherwise,
correspond to $\lambda=15$ and $w=1$.

\subsection{Clustering of Floquet Eigenstates}
\label{hsf}

In this subsection, we discuss the structure of the Hilbert space
spanned by Floquet eigenstates of the driven staggered Rydberg
chain. To this end, we plot the half-chain entanglement entropy,
$S_{L/2}$, of these eigenstates as a function of their quasienergy
$E_F$. The details of the numerical procedure used for this
computation is same as that charted in Ref.\ \onlinecite{ethv1}. The
first step involves writing down the density matrix $\rho_n =
|\chi_n\rangle\langle \chi_n|$ corresponding to the $n^{\rm th}$
Floquet eigenstate. One then integrates out the contribution to this
state from Fock states residing on one half of the chain; such an
operation has to be carried out numerically without violating the
constraint condition as detailed in Ref.\ \onlinecite{ethv1}. This
leads to the reduced density matrix $\rho_n^{\rm red}$. The
von-Neumann entanglement entropy is then computed from $\rho^{\rm
red}_n$ following the standard prescription: $S_{L/2}^{(n)} \equiv
S_{L/2} = -{\rm Tr} [\rho_n^{\rm red} \ln \rho_n^{\rm red}]$. For
finite-sized periodically driven systems, the prediction of ETH for
$S_{L/2}$ is given by the Page formula: $S_{L/2}^{\rm Page}= \ln{\rm
HSD} -1/2$, where ${\rm HSD}$ denotes Hilbert space dimension of the
half-chain with open boundary condition, and ${\rm HSD}=377$ for
$L=24$.
\begin{widetext}
\begin{figure*}
\centering{\ing[width=0.49\linewidth]{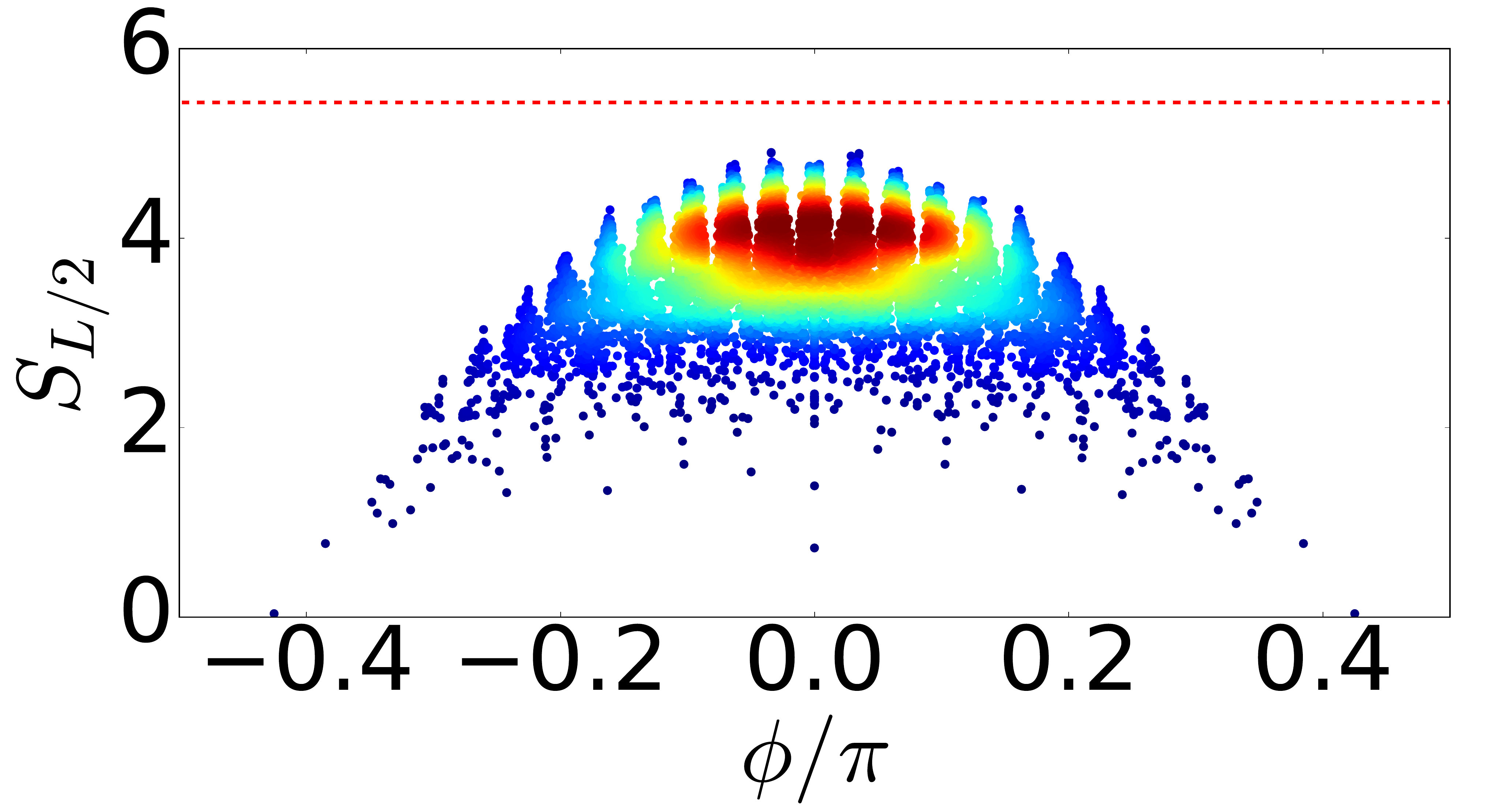}}
\centering{\ing[width=0.49\linewidth]{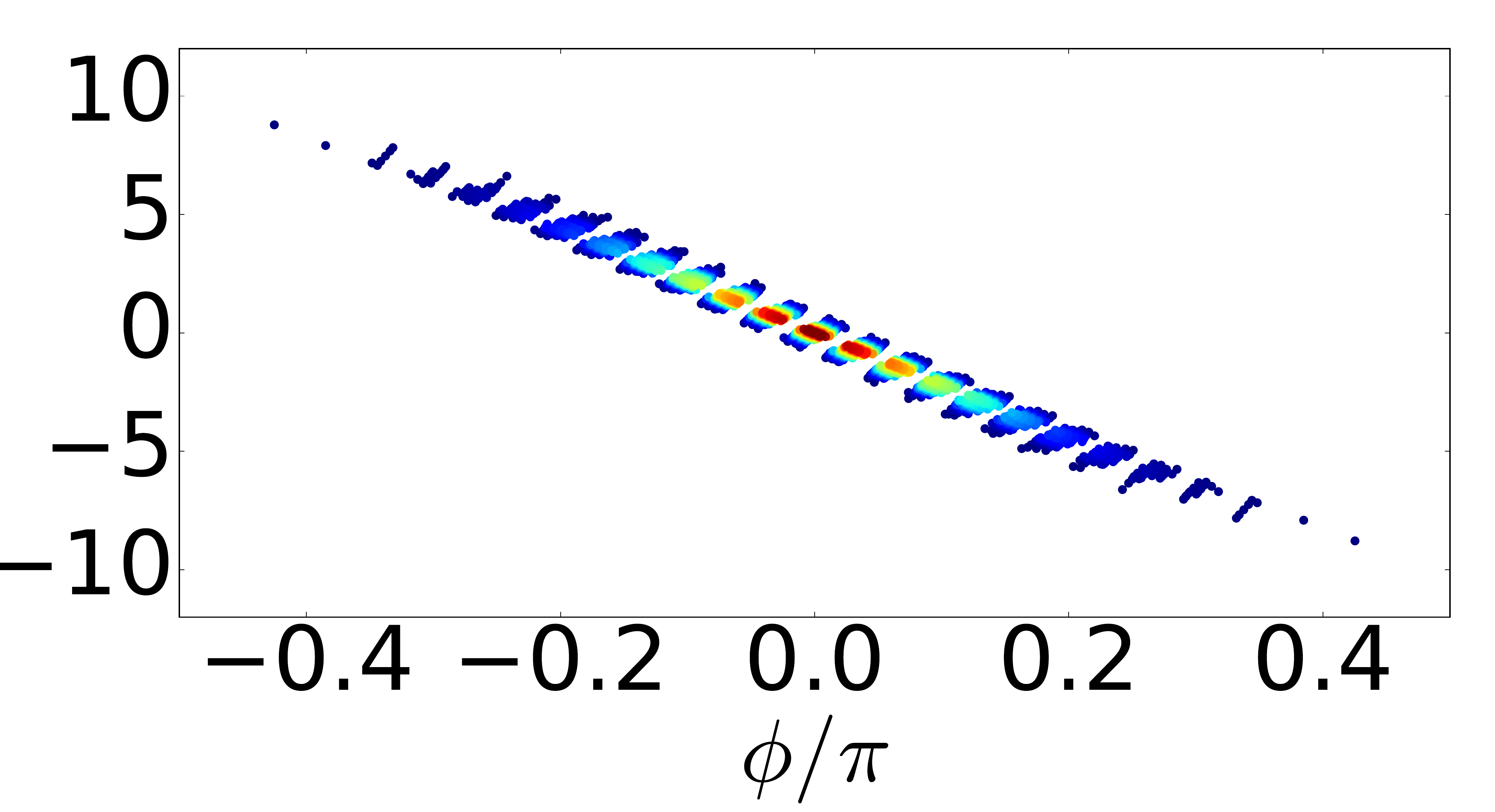}} \caption{Left:
Plot of $S_{L/2}$ as a function of $\phi=E_F T/\hbar$ for
$\omega_D=100 w/\hbar$. The red dotted line is the ETH predicted
Page value of $S_{L/2}$. Right panel: Plot of $\langle
Z_{\pi}\rangle$ for Floquet eigenstates as a function of $\phi$ for
same $\omega_D$. For both plots $L=24$, $\Delta/w=1.2$, and
$\lambda/w=15$. All energies are scaled in units of $w$ and
frequencies in units of $w/\hbar$. The color scheme indicates higher
density of states for warmer colors. See text for details. }
\label{fig1}
\end{figure*}
\end{widetext}

\begin{widetext}
\begin{figure*}
\centering{\ing[width=0.49\linewidth]{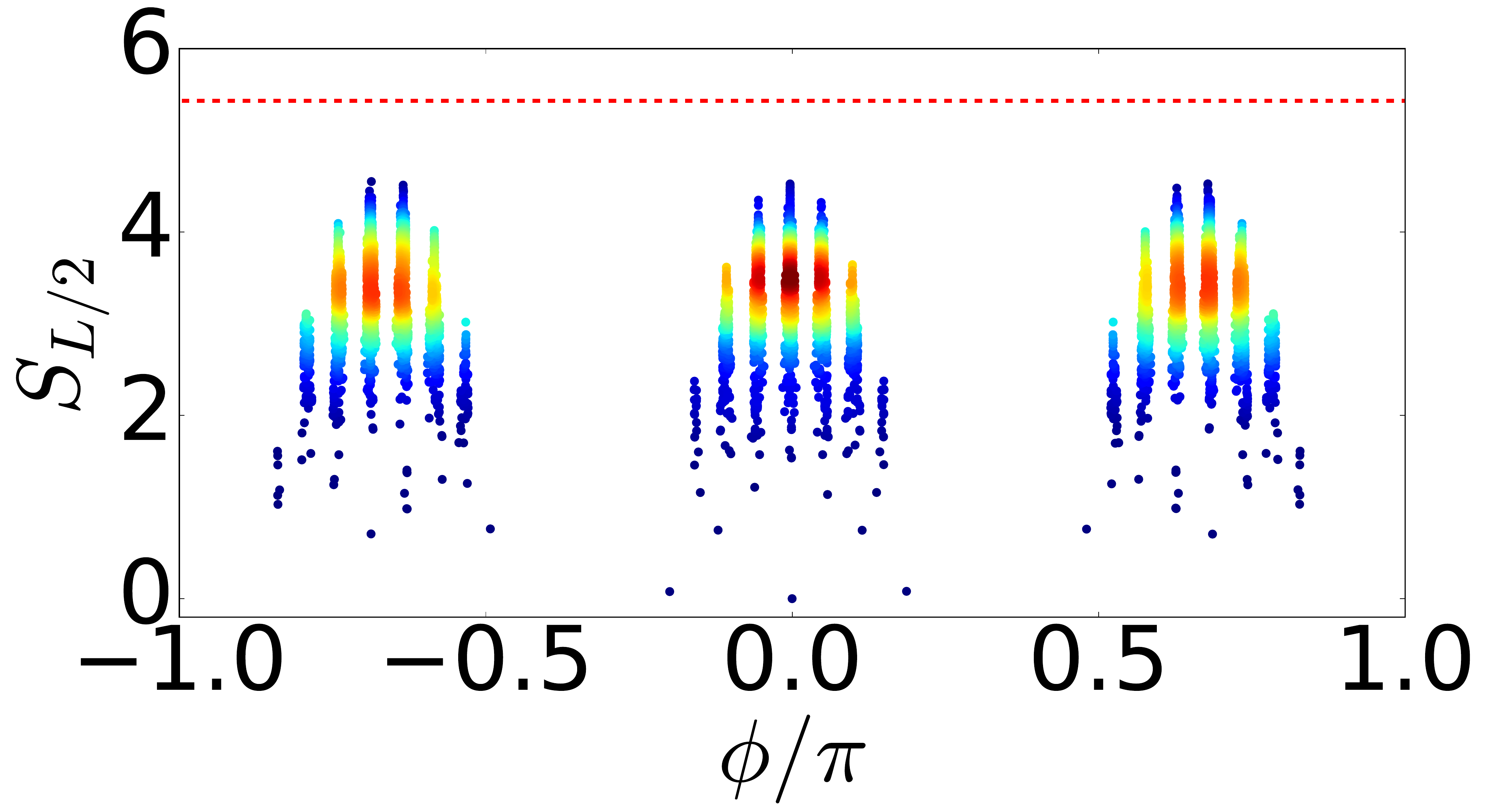}}
\centering{\ing[width=0.49\linewidth]{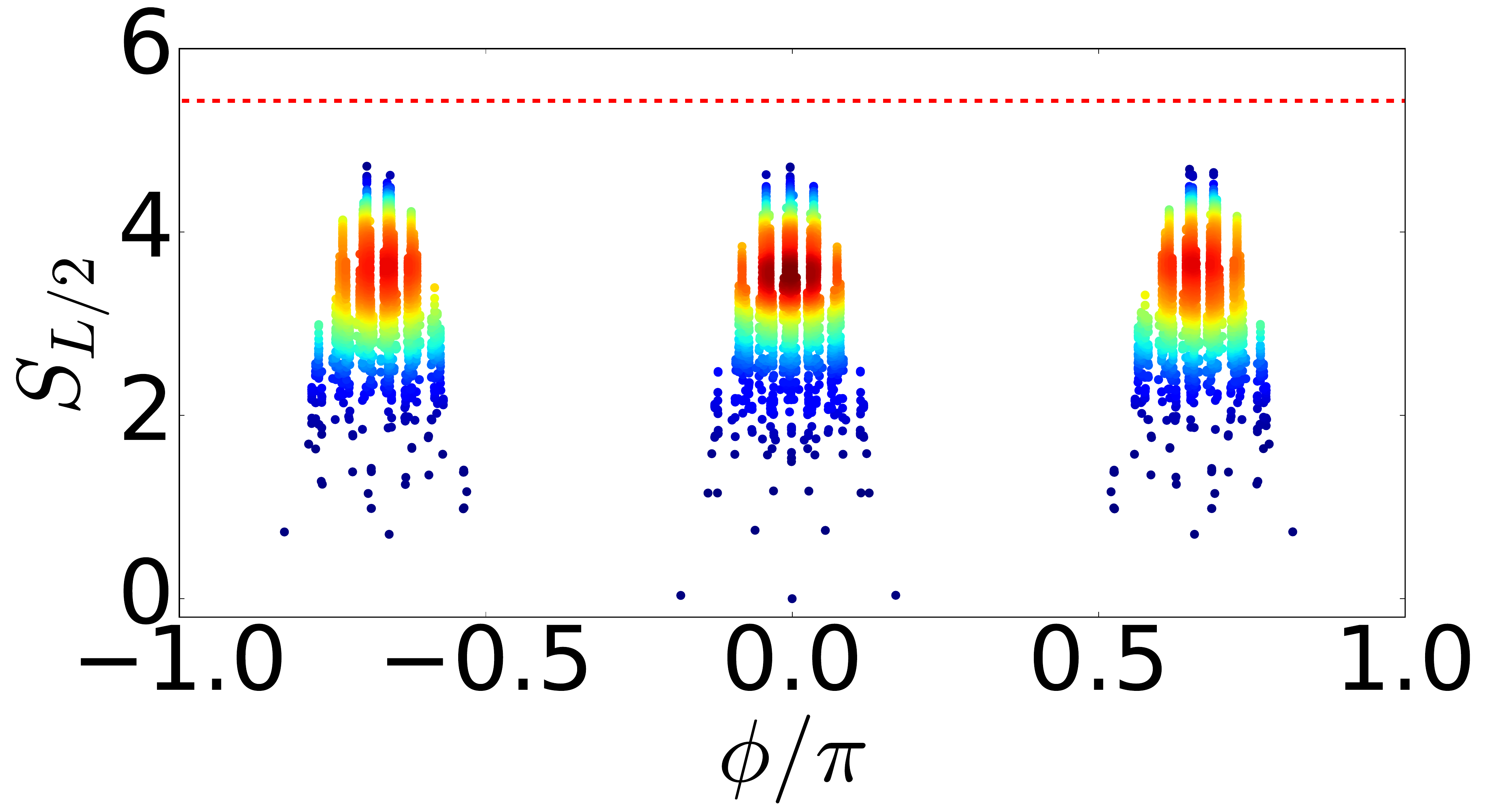}}
\centering{\ing[width=0.49\linewidth]{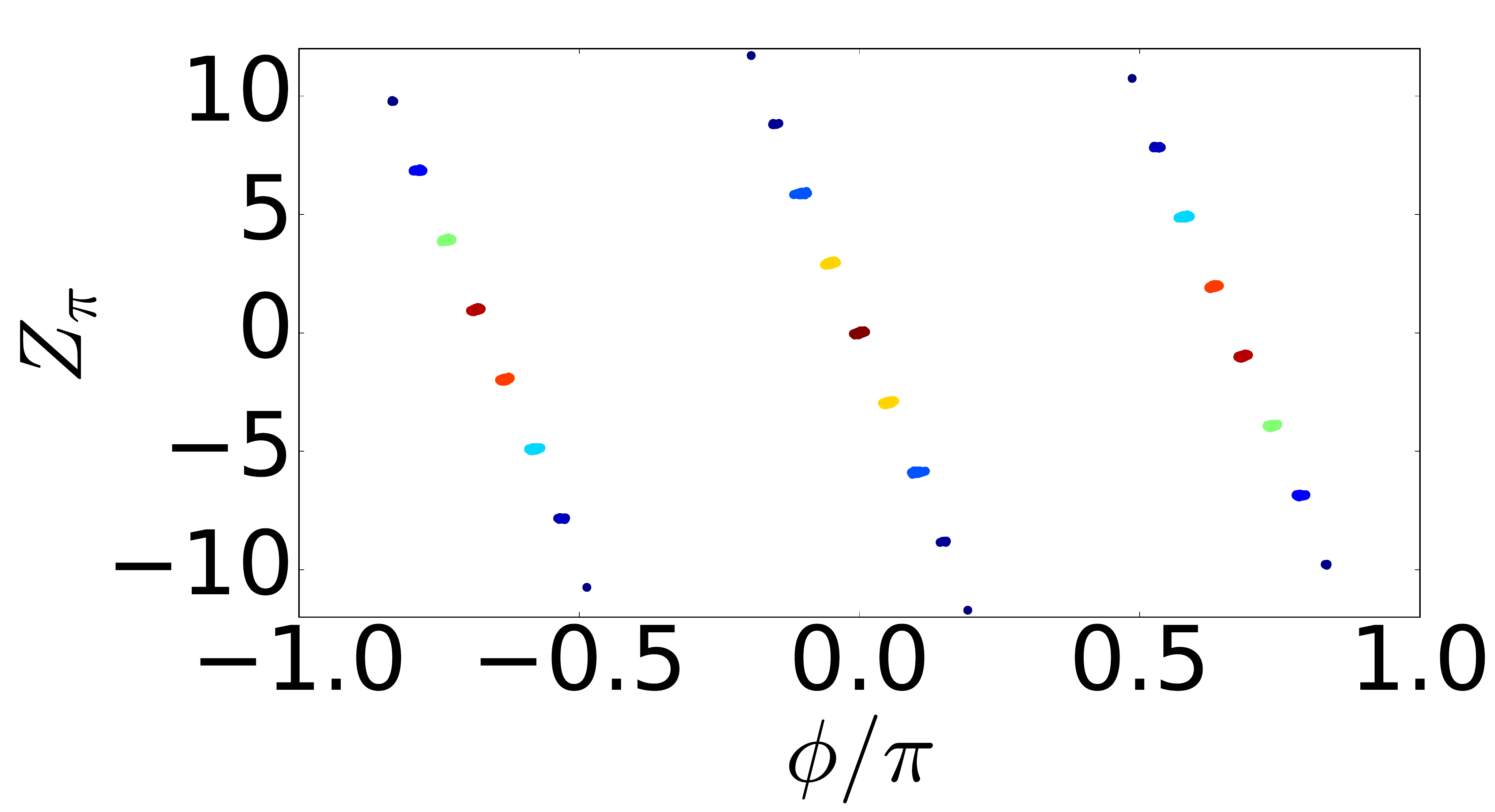}}
\centering{\ing[width=0.49\linewidth]{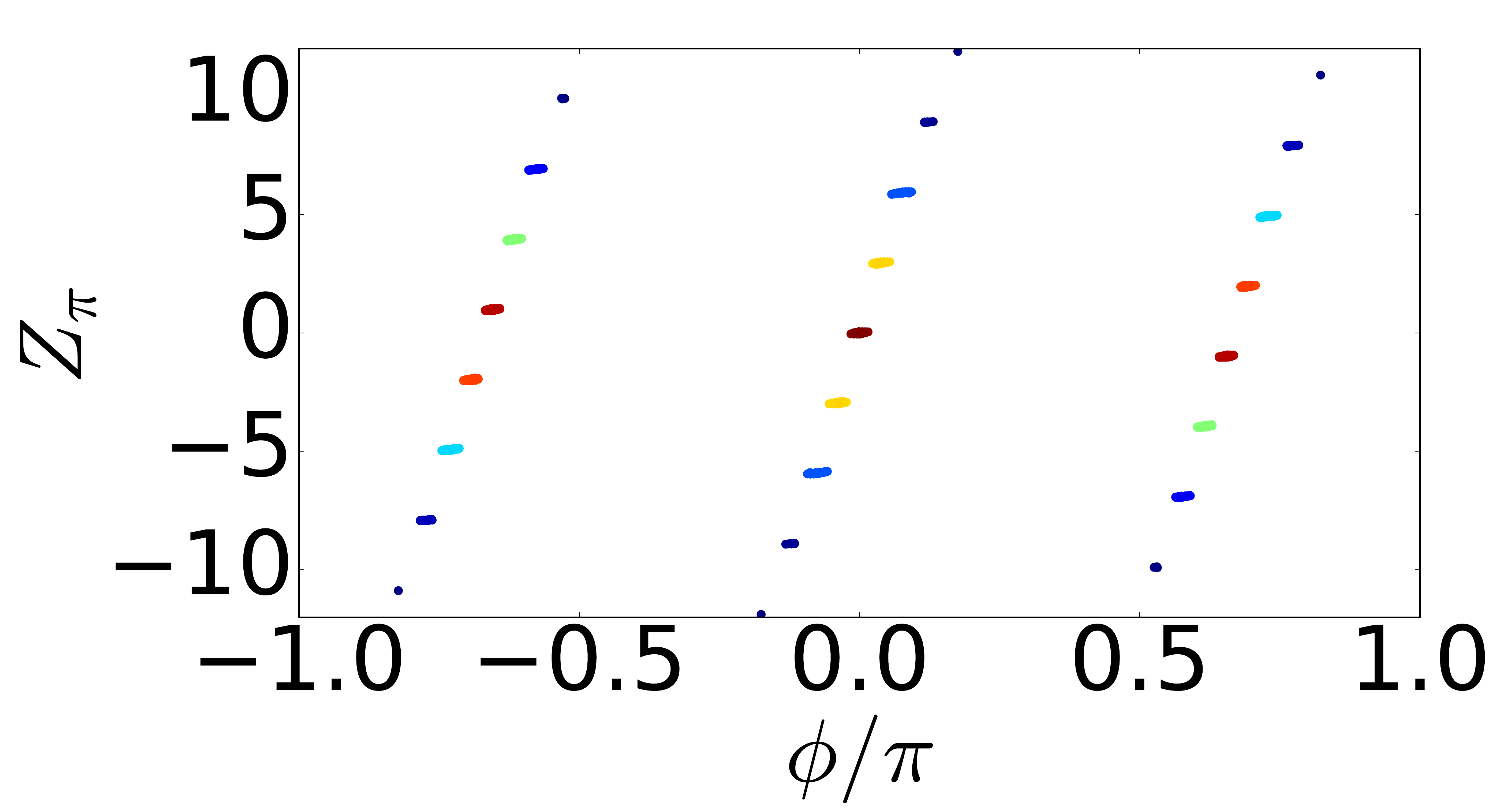}} \caption{Top left
panel: Plot of $S_{L/2}$ as a function of Floquet quasienergies
$\phi=E_F T/\hbar$ for $\omega_D=3.5 w/\hbar$. Top right panel: A
similar plot for $\omega_D=3.65 w/\hbar$. Bottom left panel: Plot of
$\langle Z_{\pi} \rangle$ for Floquet eigenstates as a function of
$\phi$ for $\omega_D=3.5 w/\hbar$. Bottom right panel: Similar plot
as the bottom left panel but for $\omega_D=3.65 w/\hbar$. For all
plots $L=24$, $\Delta=1.2 w$, and $\lambda=15 w$. All energies are
scaled in units of $w$ and frequencies in units of $w/\hbar$. The
color scheme is same as in Fig.\ \ref{fig1} and the red dotted lines
in the top panel indicate ETH predicted Page value of $S_{L/2}$. See
text for details. } \label{fig2}
\end{figure*}
\end{widetext}

The result of such a computation is shown in Figs.\ \ref{fig1} and
\ref{fig2} for $\Delta=1.2 w$. From Fig.\ \ref{fig1}, we find that
the Hilbert space is ergodic at high drive frequencies for which
$w_r \simeq w$. In contrast, as seen from Fig.\ \ref{fig2}, it
fragments into several clusters at intermediate $\omega_D$. We dub
these clusters as primary clusters. Their origin can be
straightforwardly understood as follows.

We note that the leading terms in the Floquet Hamiltonian given by
$H_{F1}$. For $\lambda T/\hbar \le 1$ ($\omega_D=100 w/\hbar$), when
$w_r \sim w$, the amplitude of the staggered and the PXP terms in
$H_{F1}$ are comparable and their eigenstates span the entire
Hilbert space as expected for a non-integrable model. This behavior
is shown in the left panel of Fig.\ \ref{fig1} and is further
highlighted by plotting the expectation value of
\begin{eqnarray}
Z_{\pi}= \frac{1}{2} \sum_j (-1)^j \sigma_j^z \label{zeq}
\end{eqnarray}
for these states, as shown in the right panel of Fig.\ \ref{fig1}.
We find that the quasienergy difference between the states with
different $\langle Z_{\pi} \rangle$ are $O(w)$; hence $w_r \sim w$
can hybridize states with different $\langle Z_{\pi}\rangle$ leading
to ergodicity.

In contrast, for $\omega_D \sim 3.5 w/\hbar$ and $3.65 w/\hbar$,
$w_r \ll w, \Delta$. In this limit, the model is near-integrable and
the eigenstates cluster into groups as shown in the top panels of
Fig.\ \ref{fig2}. Each of these group have states with definite
values $\langle Z_{\pi}\rangle$ and dimensionless quasienergy
\begin{eqnarray}
\phi &=&  \frac{E_F T}{\hbar} \simeq - \frac{\Delta T}{\hbar}
\langle Z_{\pi} \rangle \label{dimq}
\end{eqnarray}
as shown in the bottom panels panel of Fig.\ \ref{fig2}. It is
crucial to note here that $\phi$ is $2\pi$ periodic; thus Floquet
eigenstates with quasienergy outside the range $-\pi \le \phi \le
\pi$ have to be folded back, using this periodicity, to the first
Floquet-Brillouin zone. The primary clustering is therefore a
consequence of near-integrability of $H_F$ for small $w_r/\Delta$.

The eigenstates within the central primary cluster seen in the top
panel of Fig.\ \ref{fig2} has $\langle Z_{\pi} \rangle= \pm 3n$
while the other clusters have $\langle Z_{\pi}\rangle= \pm 3n + 1$
and $\pm 3n +2$ where $n$ is an integer. The number of such primary
clusters can be deduced from the fact that for the parameter regime
of these plots $\Delta/(\hbar \omega_D) \sim 1/3$. Importantly,
$\Delta$ is still incommensurate to the drive frequency by a small
amount (the exact commensuration occurs at $\Delta = \omega_D/3 \sim
1.2 w$) so that states within the same cluster are non-degenerate.
This incommensuration can be parameterized by a dimensionless ratio
given by
\begin{eqnarray}
x= \frac{\hbar \omega_D}{\Delta} -3.  \label{xdef}
\end{eqnarray}
For $\Delta=1.2$, $x\simeq -0.0834$ for $\omega_D=3.5 w/\hbar$ and
$0.04167$ for $\omega_D=3.65 w/\hbar$. The primary clusters, for
$w_r \ll \Delta$, are separated with quasienergies $\phi_1$ while
the states within each cluster are separated by smaller quasienergy
scale $\phi_2$, where
\begin{eqnarray}
\phi_1 &=& \delta E_1 T/\hbar \sim 2\pi /(3+x) \nonumber\\
\phi_2 &=& \delta E_2 T/\hbar \sim |x| 2\pi /(3+x). \label{prsedef}
\end{eqnarray}
In the parameter regime of our numerics, where $w_r \ll \delta E_1$,
the primary clustering is not destroyed by $w_r$ for finite sized
chains with $L\le 24$.

\begin{widetext}
\begin{figure*}
\centering{\ing[width=0.49\linewidth]{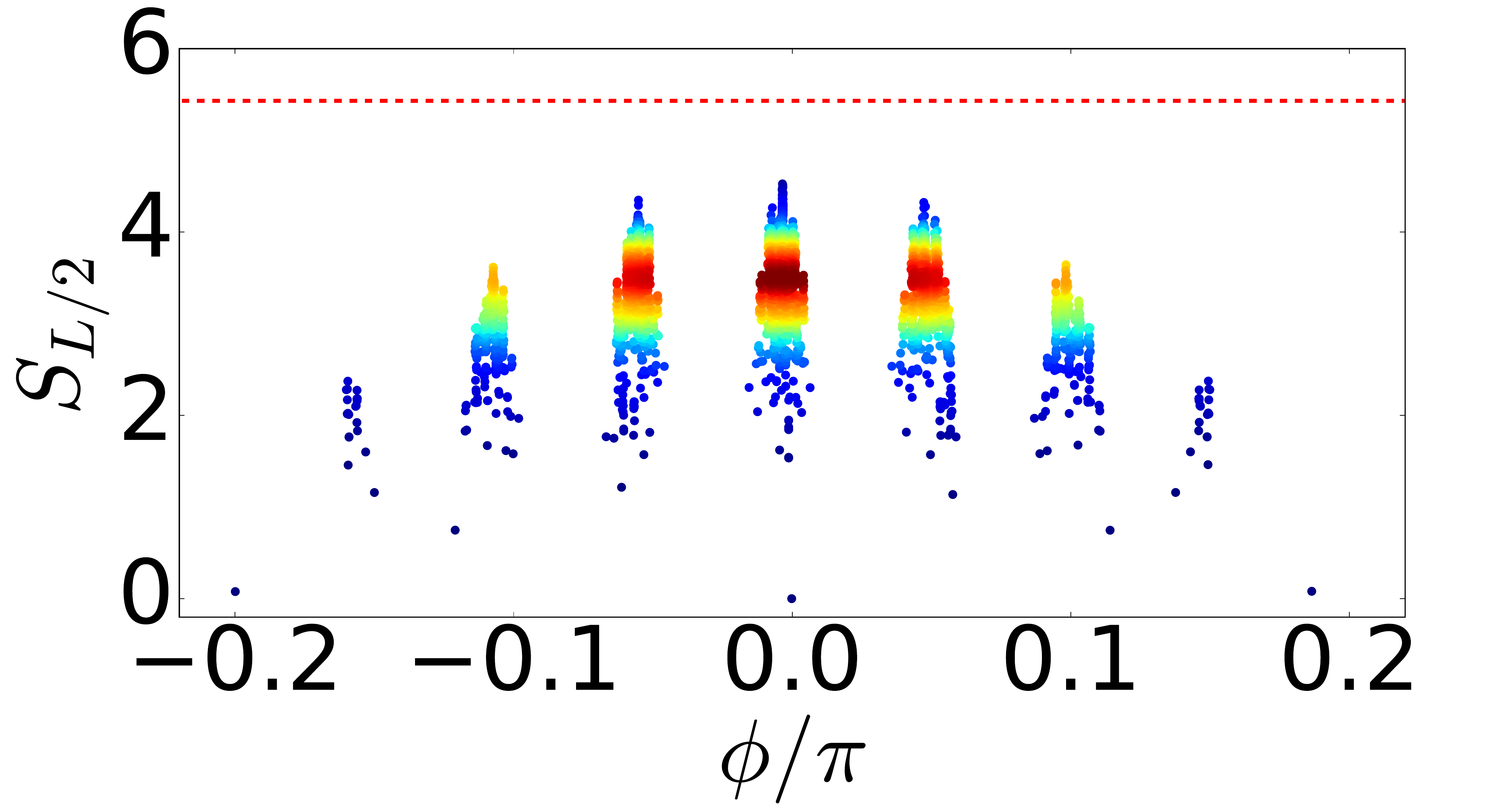}}
\centering{\ing[width=0.49\linewidth]{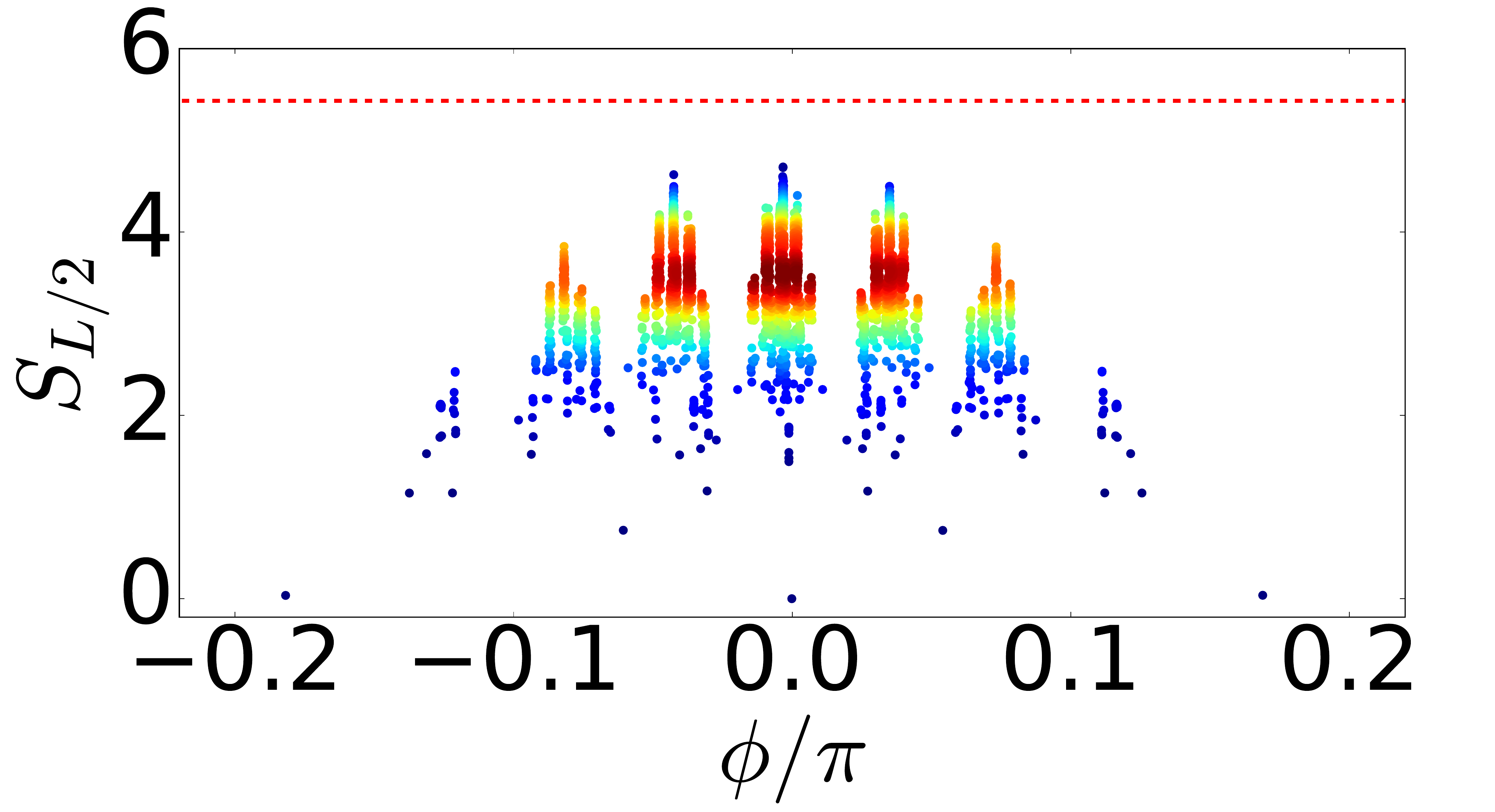}} \caption{Left
panel: Plot of $S_{L/2}$ as a function of Floquet quasienergies
$\phi=E_F T/\hbar$ within the central primary cluster for
$\omega_D=3.5 w/\hbar$. Top right panel: A similar plot for
$\omega_D=3.65 w/\hbar$. For all plots $L=24$, $\Delta=1.2 w$, and
$\lambda=15 w$. All energies are scaled in units of $w$ and
frequencies in units of $w/\hbar$. The color scheme is same as in
Fig.\ \ref{fig1} and the red dotted lines in the top panel indicate
ETH predicted Page value of $S_{L/2}$. See text for details. }
\label{fig3}
\end{figure*}
\end{widetext}

\begin{figure}
\centering{\ing[width=\linewidth]{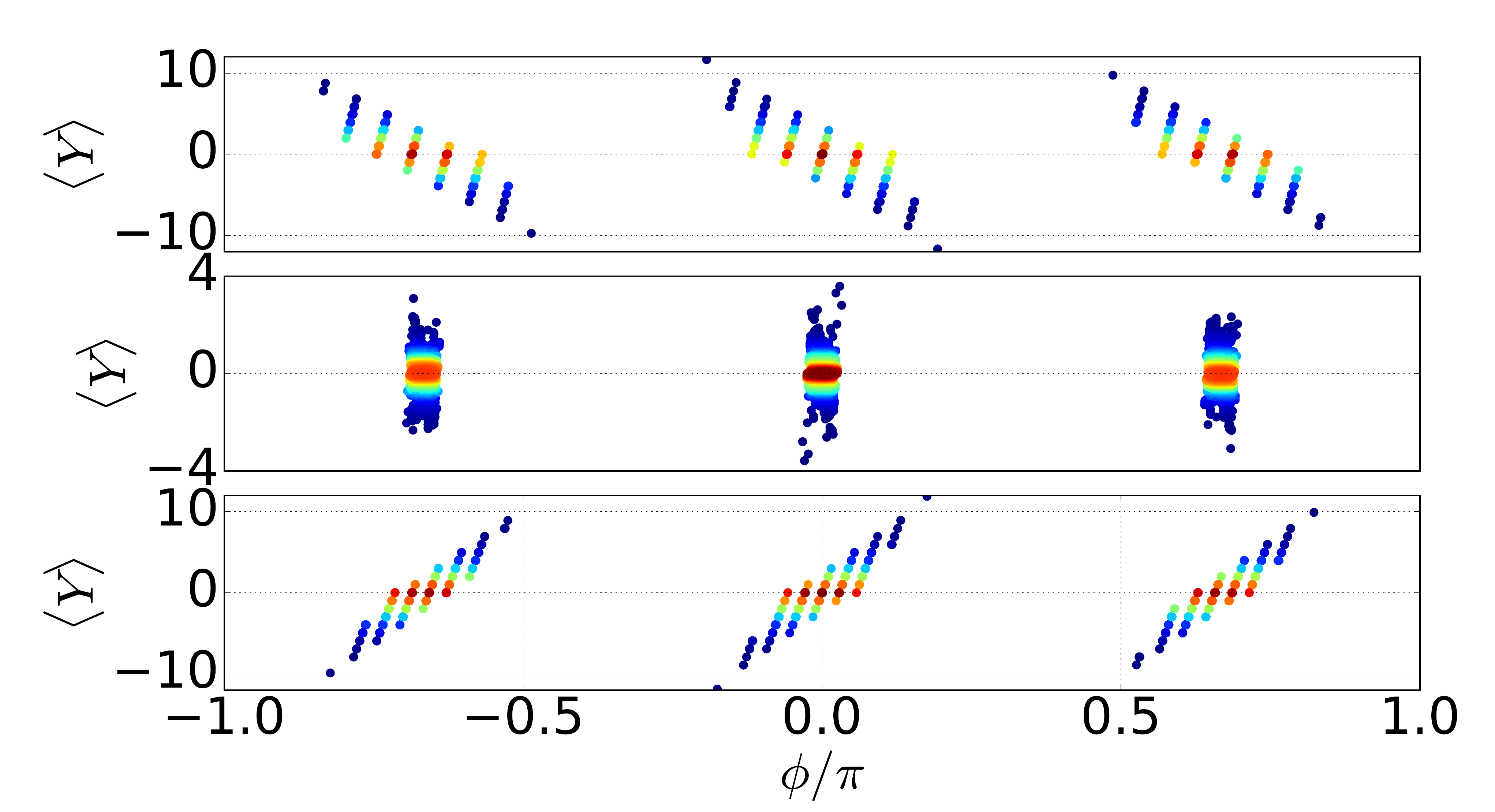}} \caption{Plot of
$\langle Y\rangle$ for the Floquet eigenstates as a function of
$\phi= E_F T/\hbar$ for $\hbar \omega_D=3.65 w$ (top panel), $3.58
w$ (middle panel) and $3.5 w$ (bottom panel). For all plots $L=24$,
$\lambda/w=15$, and $\Delta/w=1.2$. All energies are scaled in units
of $w$ and frequencies in units of $w/\hbar$. The color schme is
same as in Fig.\ \ref{fig1} and the black dashed lines are guide to
the eye. See text for details. } \label{figy}
\end{figure}

A closer inspection of the primary clusters reveal that each of
them, with definite values of $Z_{\pi}$, exhibit further clustering.
The left[right] panel of Fig.\ \ref{fig3} shows such clustering for
$\omega_D=3.5 [3.65] w/\hbar$. We dub this phenomenon as secondary
clustering. The origin of such secondary clustering can be
understood via construction of an effective Floquet Hamiltonian in
the regime $w_r \sim \delta E_2 \ll \delta E_1$. The analysis is
same as what has been carried out in Ref.\ \onlinecite{ethv05b} and
have been detailed in the appendix. Such a Hamiltonian, given by
Eqs.\ \ref{pertapp3} and \ref{pertapp4}, shows that $w_r$ can have
significant non-zero matrix elements only between states within each
primary cluster. Moreover, as shown in Ref.\ \onlinecite{ethv05b},
the presence of a finite $w_r$ does not lead to ergodic structure
within each primary cluster. Instead, a perturbative analysis,
similar to that carried out in Ref.\ \onlinecite{ethv05b} and
detailed in the appendix, indicates that the secondary clustering
occurs due to an emergent conserved quantity (Eq.\ \ref{pertapp4})
\begin{eqnarray}
Y= -\sum_j (-1)^j P_{j-1} \sigma_j^z P_{j+1} \label{yexp}
\end{eqnarray}
within each $Z_{\pi}$ sector. These secondary clusters correspond to
groups of states with same values of $\langle Y \rangle$. This can
be clearly seen from Fig.\ \ref{figy} where $\langle Y\rangle$ is
plotted for eigenstates in each of the primary cluster. We find a
definite value of $\langle Y\rangle$ for all Floquet eigenstates
both at $\hbar \omega_D=3.65 w$ and $3.5 w$ as can be seen from the
top and bottom panels of Fig.\ \ref{figy} respectively. We note that
the emergence of $Y$ relies on the perturbative parameter
$w_r/\delta E_2 \le 1$. Such a perturbation theory will break down
for at the exact commensuration point where $x$, and hence $\delta
E_2$, vanishes. This leads to loss of secondary clustering. At these
commensuration points, the Floquet eigenstates can not be associated
with a definite $\langle Y\rangle$ value as can be seen from the
middle panel of Fig.\ \ref{figy}. We shall discuss this case in
Sec.\ \ref{ergres}.

\subsection{Ergodicity restoration} \label{ergres}

\begin{widetext}
\begin{figure*}
\centering{\ing[width=0.3\linewidth]{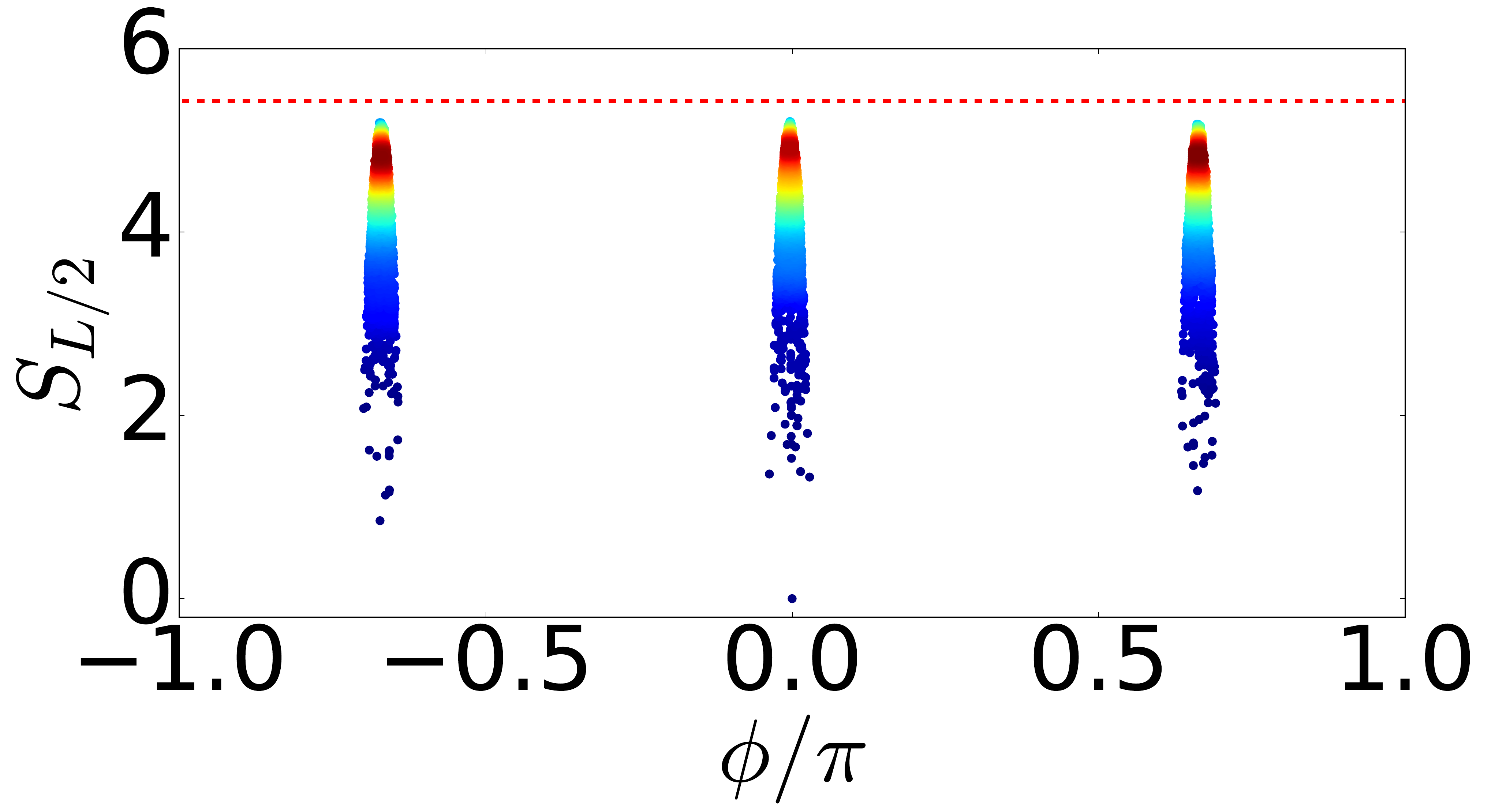}}
\centering{\ing[width=0.3\linewidth]{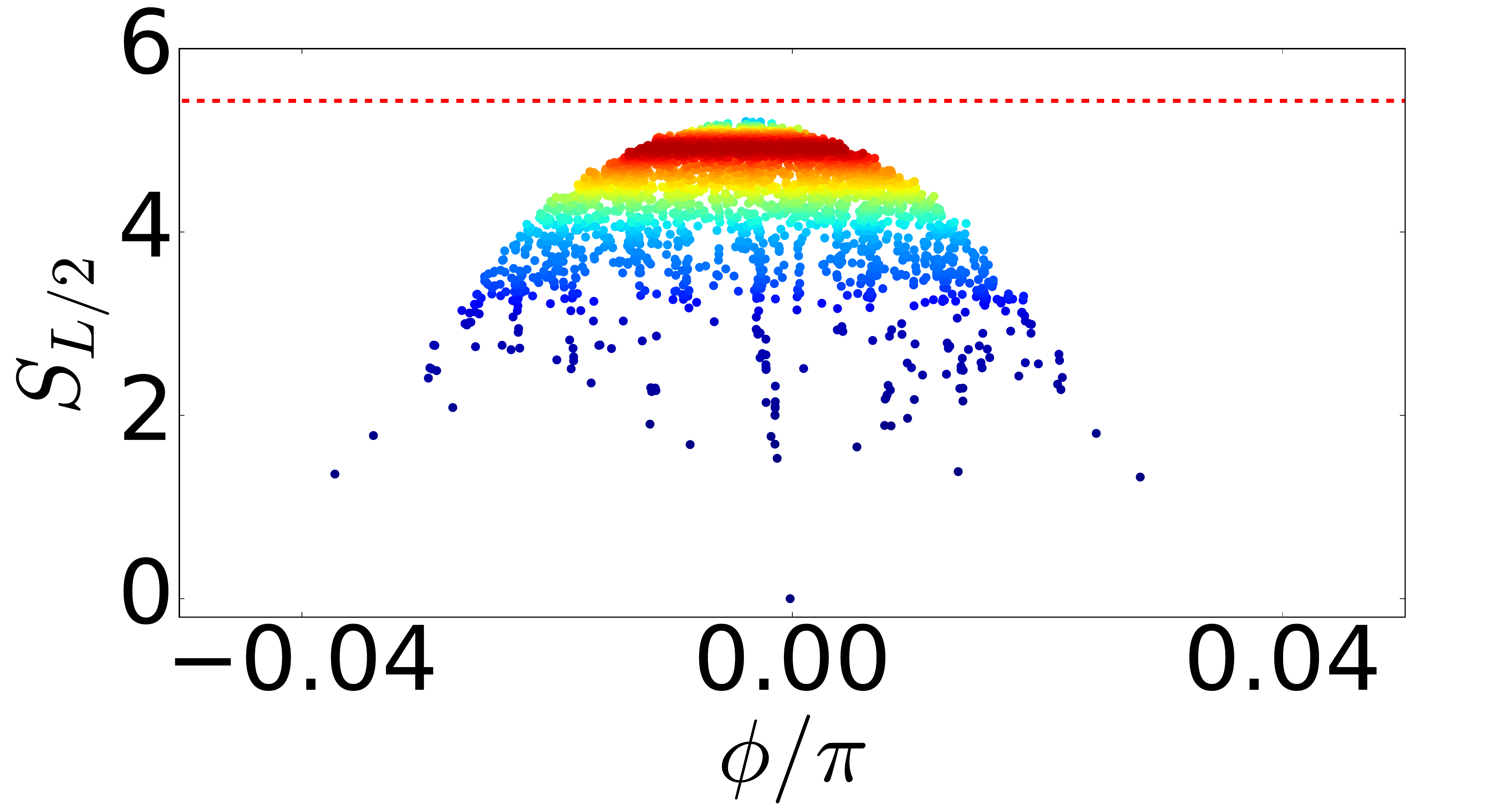}}
\centering{\ing[width=0.3\linewidth]{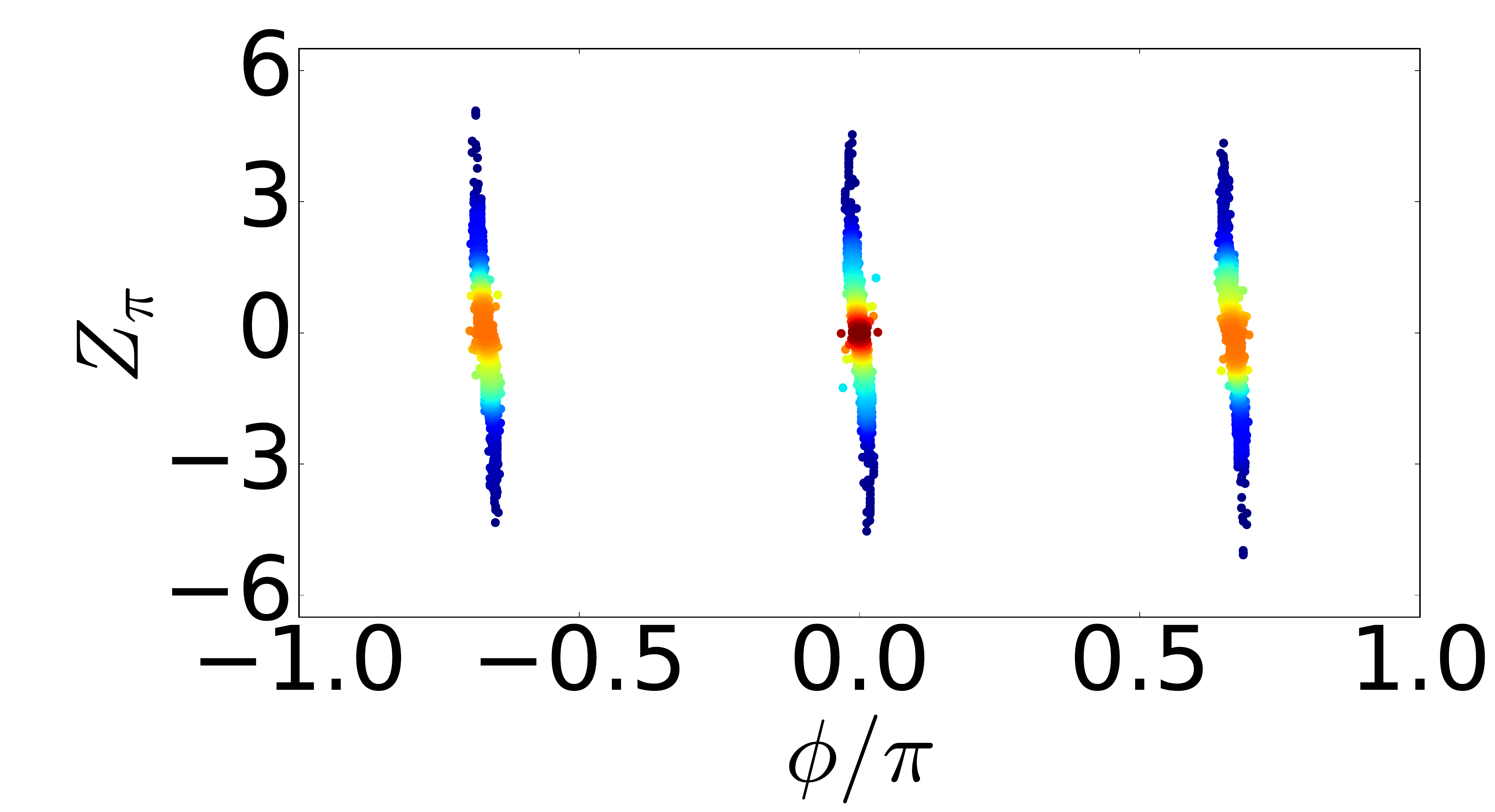}}\caption{Left panel:
Plot of $S_{L/2}$ as a function of $\phi= E_F T/\hbar$ for
$\omega_D=3.58 w/\hbar$. Central panel: A similar plot showing
states within the primary cluster around $E_F=0$. Right Panel: Plot
of $\langle Z_{\pi} \rangle$ as a function of $\phi$. For all plots
$\Delta/w=1.2$, and $\lambda/w=15$. All energies are scaled in units
of $w$ and frequencies in units of $w/\hbar$. The color scheme is
same as in Fig.\ \ref{fig1} and the red dotted lines in the left and
central panels indicate ETH predicted Page value of $S_{L/2}$. See
text for details. } \label{figsp1}
\end{figure*}
\end{widetext}

\begin{figure}
\centering{\ing[width=\linewidth]{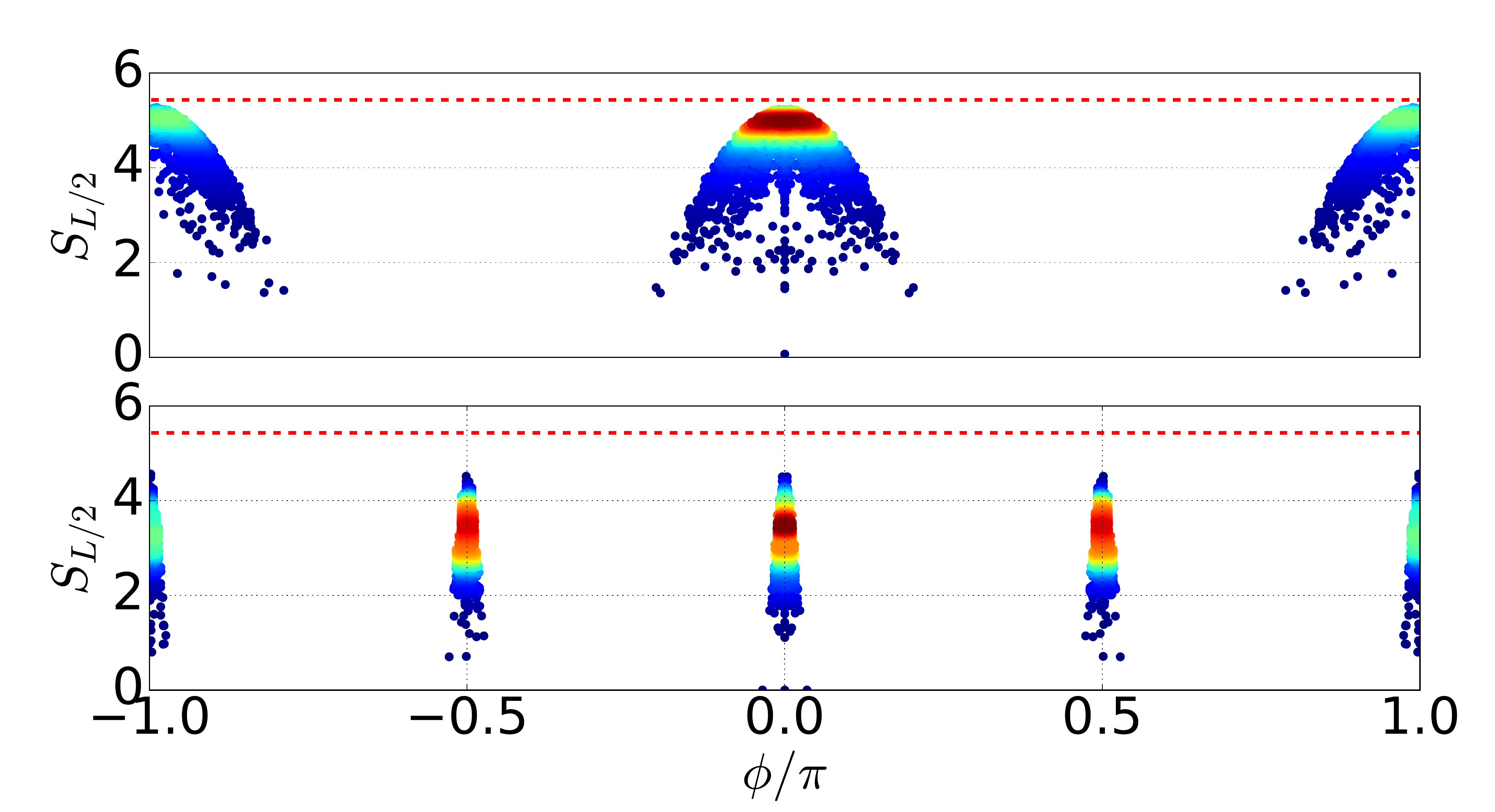}} \caption{Top Panel:
Plot of $S_{L/2}$ as a function of Floquet quasienergies $\phi=E_F
T/\hbar$ for $\Delta/w = 1.8$ showing two primary clusters. Bottom
panel: A similar plot for $\Delta/w =0.9$ showing four primary
clusters. For both plots $\hbar \omega_D/w=3.6 w$, and
$\lambda/w=15$. All energies are scaled in units of $w$ and
frequencies in units of $w/\hbar$. The color scheme is same as in
Fig.\ \ref{fig1} and the red dotted lines indicate ETH predicted
Page value of $S_{L/2}$. The black dashed lines are guide to the
eye. See text for details. } \label{figfrag}
\end{figure}

In this section, we discuss the behavior of the driven chain when
the amplitude of the staggered detuning is commensurate with the
drive frequency. To this end, we note that the Floquet Hamiltonian
derived using FPT in Sec.\ \ref{fpt} can be written as
\begin{eqnarray}
H_F &=& - \sum_j \Delta_r Z_{\pi} + H'_1 \nonumber\\
H'_1 &=& -w_r \sum_j \tilde \sigma_j^x  + H_{F3}^{(2)} + H_F^{(3)} +
..  \nonumber\\
\Delta_r &=& \Delta \left[ 1- \frac{w^2 T^2}{8 \gamma^3}
\left(\frac{1}{2} \sin 2 \gamma - \gamma \cos 2 \gamma\right) +...
\right] \label{fptform}
\end{eqnarray}
where the ellipsis indicate higher order contribution to the Floquet
Hamiltonian and $\Delta_r$ denote the renormalized amplitude of the
staggered term. We note that for $\hbar \omega_D/\Delta_r = n$ where
$n \in Z$, one can write
\begin{eqnarray}
U(T,0) &=& e^{- i H_F T/\hbar} = e^{-i (H'_1 T +(2 \pi/n) Z_{\pi})
/\hbar} \label{resfl}
\end{eqnarray}
Thus the Floquet eigenstates can be organized into clusters of
states with values of $\langle Z_{\pi} \rangle  = \pm 3n$, $\pm
3n+1$ and $\pm 3n+2$. This is seen in the top left panel of Fig.\
\ref{figsp1}. We also note that each of the states in a given
primary cluster shall have identical quasienergy at $w_r=0$ since
$x$, and hence $\delta E_2$, vanishes. Thus in absence of $w_r$,
they form a flat band and the Hilbert space comprises of three such
flat bands with $\delta E_1= 0$ and $\pm \hbar \omega_D/3$ in this
limit. The number of such clusters depends on $n$; this is exhibited
in Fig.\ \ref{figfrag} where the structure of the states with $n=2$
and $n=4$ primary clusters are shown.

In the presence of a non-zero $w_r$, within each cluster, the
effective Floquet Hamiltonian is given by $H'_1$. For small $w_r$,
the primary clustering still exists for finite $L$, since $w_r \ll
\Delta$. However, within each primary cluster, $H_F^{\rm eff} \simeq
H'_1$ since $\delta E_2 =0$. Thus the effective Floquet Hamiltonian
for each primary cluster resembles a renormalized PXP Hamiltonian
along with higher order terms. The states within each primary
cluster are therefore expected to have a ergodic structure with no
secondary clustering. Moreover, as seen from the bottom panel of
Fig.\ \ref{figsp1}, they possess a continuous range of $\langle
Z_{\pi}\rangle$ values for any finite $w_r$. This property stems
from strong hybridization between states with definite integer
values of $\langle Z_{\pi}\rangle$ in the presence of non-zero
$w_r$. We note that the perturbative analysis of derived in Ref.\
\onlinecite{ethv05b} leading to the emergent invariant $Y$
responsible for the secondary clustering becomes invalid here since
the perturbative parameter $w_r/\delta E_2 \to \infty$ at the
commensuration point. This expectation is validated from exact
numerics as seen from the top right panel of Fig.\ \ref{figsp1}
where the ergodic structure for states within the central primary
cluster is shown. We note that each of the clusters contain both
anomalous states with low $S_{L/2}$ along with thermal states with
high $S_{L/2}$. A similar behavior is expected for the $s=\pm 1$
primary clusters.

\begin{figure}
\centering{\ing[width=\linewidth]{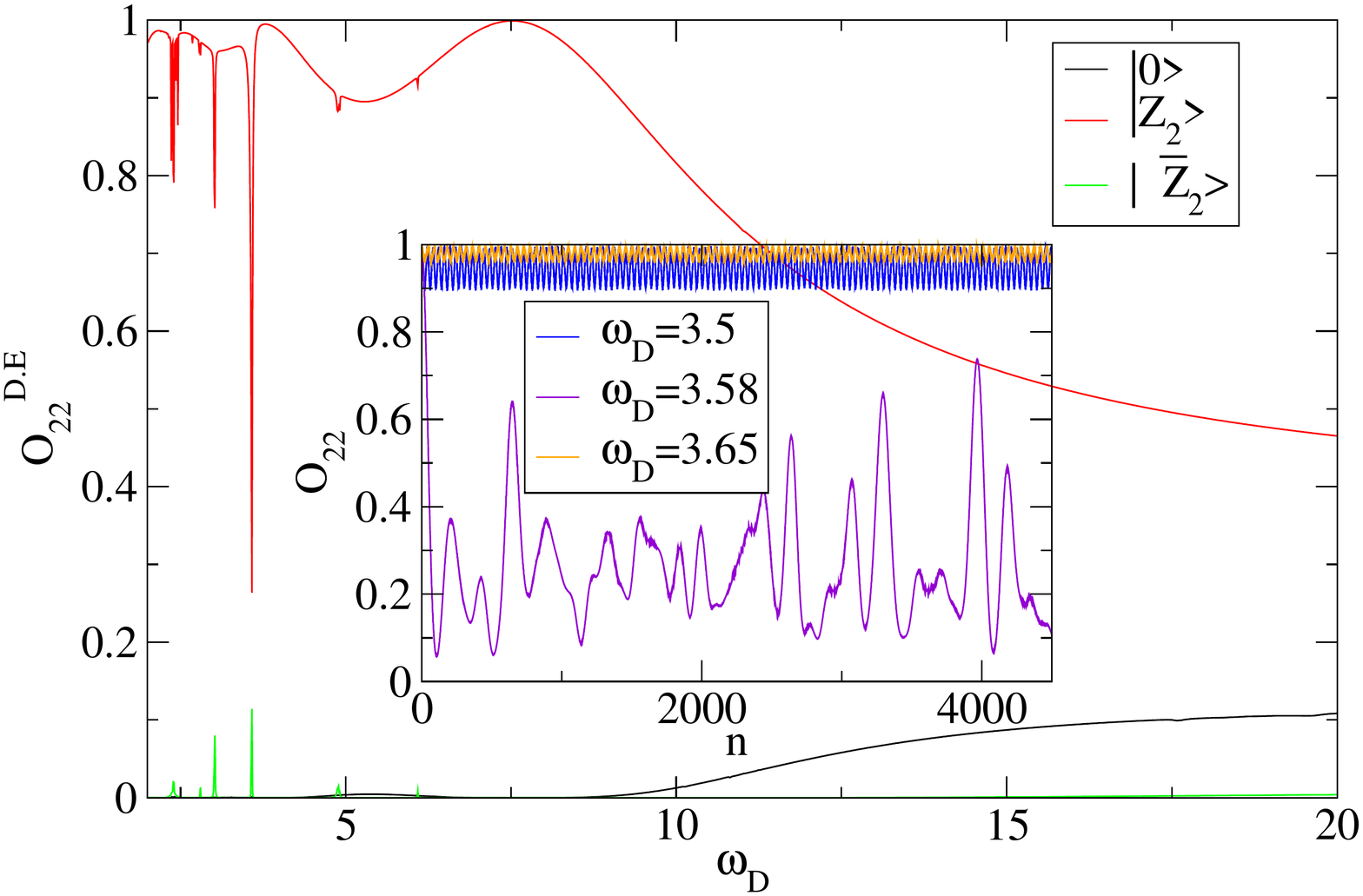}} \caption{Plot of the
steady state value of $O_{22} \equiv O_{22}^{\rm DE}$ as a function
$\omega_D$ for $\Delta/w=1.2$ showing restoration of ergodicity at
special frequencies. Such restoration is characterized by sharp dips
in the $O_{22}^{\rm DE}$ value. The inset shows the dynamics of
$O_{22}$ starting from a $|{\mathbb Z}_2\rangle$ state as a function
of $n$ for commensurate ($\hbar\omega_D/w=3.58$) and slightly
incommensurate ($\hbar\omega_D/w=3.65$ and $\hbar \omega_D/w=3.5$)
drive frequencies. The coherent oscillations of $O_{22}$ (shown
above for $\hbar \omega_d/w=3.5$ and $3.65$) seen at incommensurate
frequencies are replaced by a slower approach to the steady state at
exactly commensurate drive frequency ($\hbar \omega_D/w=3.58$). All
energies (frequencies) are scaled in units of $w$($w/\hbar$). See
text for details. } \label{figspdyn}
\end{figure}

\begin{figure}
\centering{\ing[width=\linewidth]{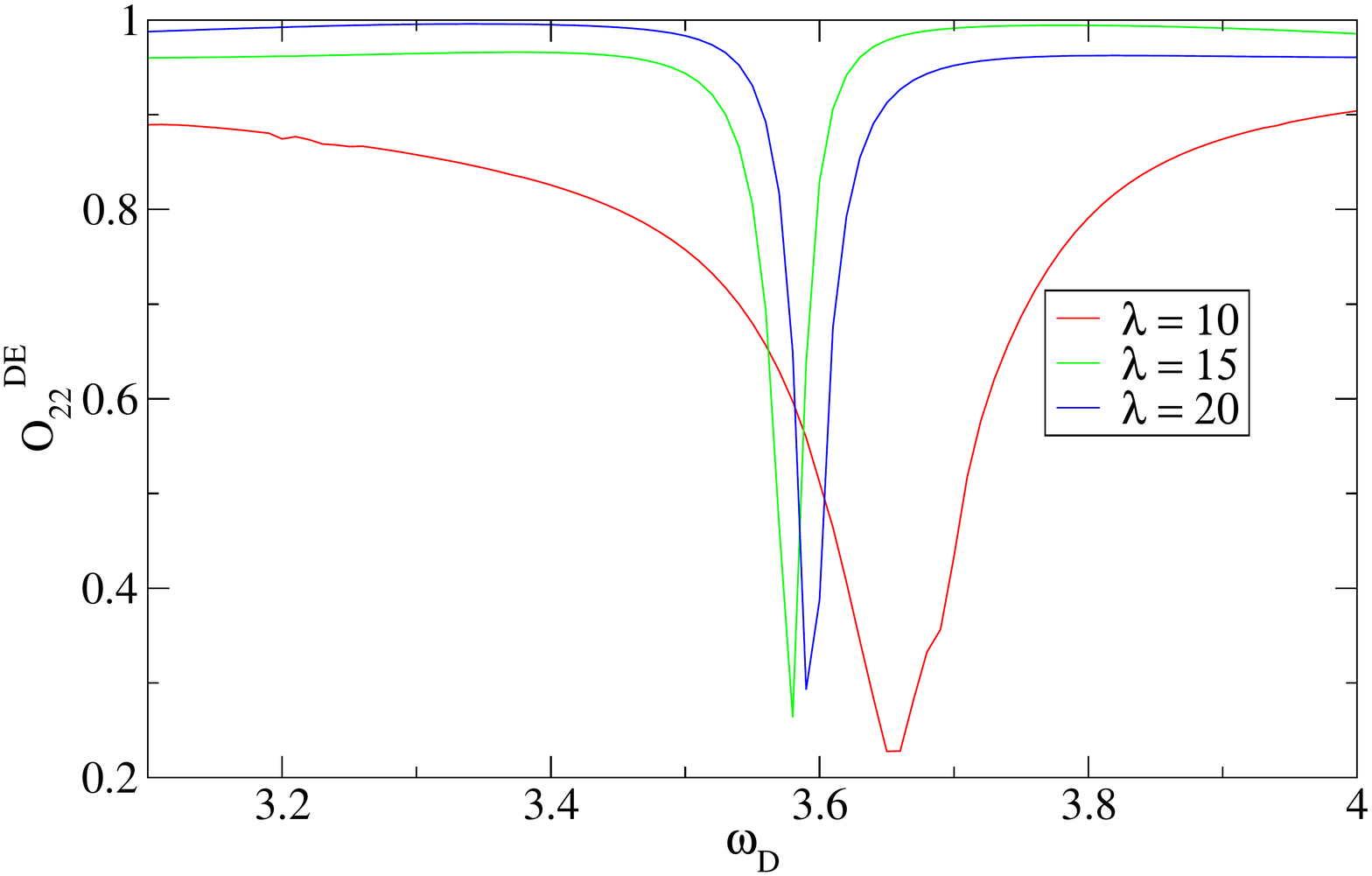}} \caption{Plot of the
steady state value of $O_{22} \equiv O_{22}^{\rm DE}$ as a function
$\omega_D$ for $\Delta/w=1.2$ for several $\lambda$ showing increase
in width of the ergodic region as a function of $\lambda$ around
$\hbar \omega_D/w= 3.58$. All energies (frequencies) are scaled in
units of $w(w/\hbar)$. See text for details. } \label{figwide}
\end{figure}

The consequence of such ergodicity restoration within a primary
cluster is also reflected in the steady state value of $O_{22}$,
$O_{22}^{\rm DE}$, as shown in Fig.\ \ref{figspdyn}. The steady
state value of $O_{22}$, in terms of the Floquet spectrum,  is given
by
\begin{eqnarray}
O_{22}^{\rm DE} &=& \sum_n |c_n^{\rm init}|^2 \langle n| \hat n_2
\hat n_4|n\rangle  \label{o22steady}
\end{eqnarray}
where $c_n^{\rm init}= \langle n|\psi_0\rangle$ and $|\psi_0 \rangle
\equiv |{\bar {\mathbb Z}_2}\rangle, \,|{\mathbb Z}_2\rangle,\, {\rm
or}\, |0\rangle$ is the overlap of the Floquet eigenstates
$|n\rangle$ with the initial state. We find that $O_{22}^{\rm DE}$
exhibits a sharp dip at the commensuration point when the dynamics
starts from $|{\mathbb Z}_2\rangle$; in contrast it exhibits a
smaller peak for $|{\bar {\mathbb Z}_2}\rangle$ initial state. In
both cases, its value approaches close to that predicted by ETH.
This is clearly a consequence of restoration of ergodicity within a
primary cluster. As one varies the frequency away from the
commensuration point, $O_{22}^{\rm DE}$ deviates sharply from its
ETH predicted value approximately around $w_r \simeq \delta E_2$.
The sharpness of such deviation is a consequence of small $w_r
\simeq 0.045$ at $\lambda=15 w$ and $\hbar \omega_D=3.58 w$. This
can be controlled by varying $\lambda$ (for a fixed $\omega_D$) as
seen in Fig.\ \ref{figwide} displaying a significantly wider ergodic
regime as a function of the drive frequency at $\lambda=5 w$ and
$\lambda=10 w$.

The stroboscopic evolution of $O_{22}$ at these ergodicity restoring
points do not display coherent oscillations as can be seen from the
inset of Fig.\ \ref{figspdyn}; instead it displays a slow approach
to the thermal steady state. We note that this is in contrast to the
dynamics of $O_{22}$ starting from $|{\mathbb Z}_2\rangle$ for a
renormalized PXP Floquet Hamiltonian. This difference can be
understood as follows. Since $w_r/\Delta \ll 1$, the dynamics of
$O_{22}$ is dominated by terms which connects states within the same
primary cluster. The matrix elements between any two such states,
which differ by at least three flipped spins for $\hbar
\omega_D/\Delta_r \simeq 3$, must involve at least ${\rm O}(w_r^3)$
term; consequently, the effective Hamiltonian for such dynamics is
different from the renormalized PXP. Thus the scar-induced
oscillations are not seen. Such an effective ${\rm O}(w_r^3)$ term
naturally leads to a thermalization time which is ${\rm
O}(w_r^{-3})$ and hence large.

\section{Dynamical freezing} \label{df}

In this section, we present our results on the density-density
correlation function $O_{j2}= \langle \hat n_j \hat n_{j+2}\rangle$
with focus on dynamical freezing of these correlations at specific
drive frequencies and their oscillations near these freezing
frequencies. For all numerical results in this section, we shall
choose $j=2$. The discussion on the freezing phenomenon is given in
Sec.\ \ref{freezsec} while the oscillation of $O_{22}$ around the
freezing point is analyzed in Sec.\ \ref{oscsec}.

\subsection{Freezing at special drive frequencies}
\label{freezsec}

\begin{figure}
\centering{\ing[width=\linewidth]{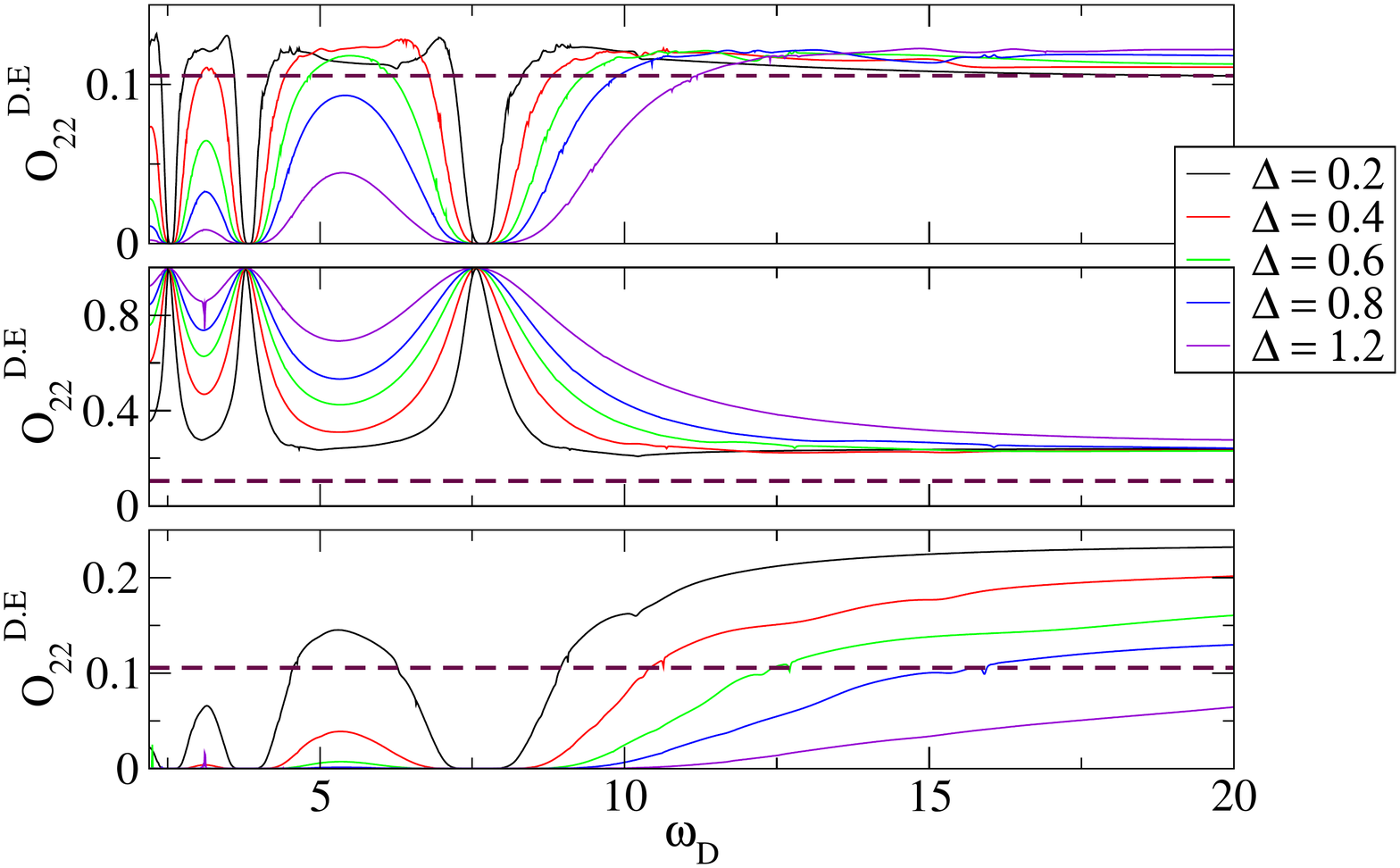}} \caption{Plot of the
steady state value of $O_{22} \equiv O_{22}^{\rm DE}$ as a function
of $\omega_D$ starting from the $|0\rangle$ (top panel), $|{\mathbb
Z}_2\rangle$ (middle panel) and $|{\bar {\mathbb Z}_2}\rangle$
(bottom panel) states for several representative values of
$\Delta/w$ as shown. The black dashed lines indicate the ETH
predicted value of $O_{22}^{\rm DE}$. For all plots $\lambda/w=15$.
All energies are scaled in units of $w$ and frequencies in units of
$w/\hbar$. See text for details. } \label{figss}
\end{figure}

We begin with a study of the behavior of $O_{22}^{\rm DE}$  as a
function of $\Delta/w$ as shown in Fig.\ \ref{figss}. The plot shows
its behavior for the initial Neel ( $|{\mathbb Z}_2\rangle$ and
$|{\bar {\mathbb Z}_2}\rangle$ ) and vacuum ($|0\rangle$) states. It
indicates strong deviation from the ETH predicted value $O_{22}^{\rm
ETH} \simeq 0.11$ at intermediate drive frequencies. For the
$|0\rangle$ initial state, $O_{22}^{\rm DE}$ reaches its ETH
predicted value at high drive frequencies where $w_r \sim w \simeq
\Delta$ as expected; in contrast, for the $|{\mathbb Z}_2\rangle$
initial state, $O_{22}^{\rm DE}$ reaches a super-thermal value. The
latter behavior can be attributed to the presence of scar states in
the Floquet spectra with large overlap with the $|{\mathbb
Z}_2\rangle$ states \cite{ethv1,ethv2}; we shall discuss these
states in Sec.\ \ref{crover}. For either of these initial states,
$O_{22}^{\rm DE}$, in the high-frequency limit, is almost
independent of $\Delta$ as long as $w \ge \Delta$. In contrast, it
depends strongly on $\Delta$ for $\hbar \omega_D \gg w$ and can take
either superthermal or subthermal values for $|{\bar {\mathbb
Z}_2}\rangle$ initial state; this constitutes a stronger violation
of ETH which can not be explained by presence of scars alone.
\begin{figure}
\centering{\ing[width=\linewidth]{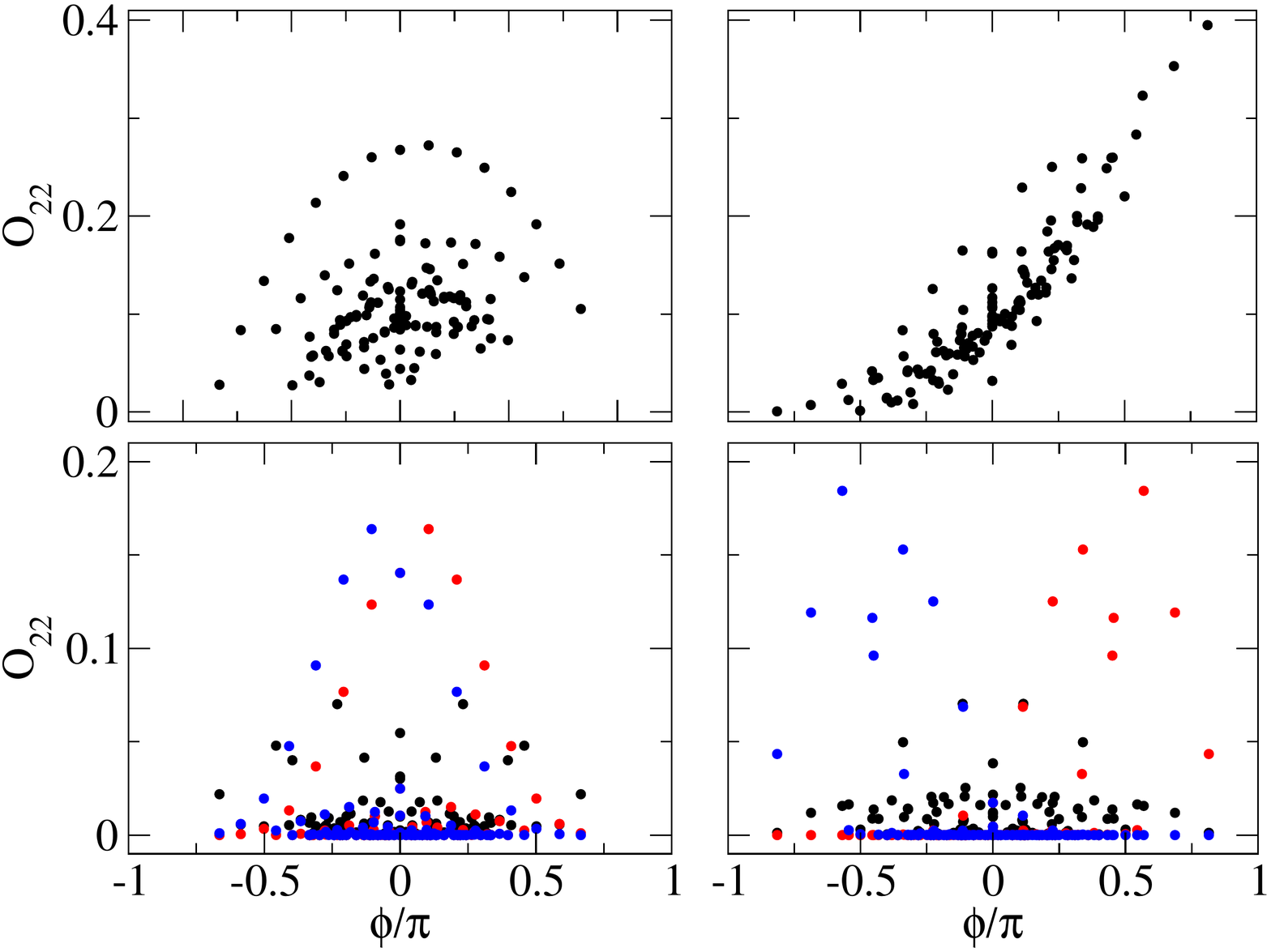}}
\caption{Top left panel: Plot of $O_{22}^{(m)}= \langle m| \hat n_2
\hat n_4|m\rangle \equiv O_{22}$ corresponding to Floquet
eigenstates $|m\rangle$ as a function of their quasienergy
$\phi=E_F^{(m)} T/\hbar$ for $\Delta/w=0.1$. Bottom left panel:
Overlap of these eigenstates with $|{\mathbb Z}_2\rangle$ (red
dots), $|{\bar {\mathbb Z}_2}\rangle$ (blue dots), and $|0\rangle$
(black dots). Right panels: Same as the corresponding left panels
but for $\Delta/w=1.2$. All left panel plots have same Y axes range
as their right panel counterparts. All plots correspond to $L=14$,
$\lambda/w=15$ and $\hbar \omega_D/w= 6.9$. All energies are scaled
in units of $w$ and frequencies in units of $w/\hbar$. See text for
details. } \label{figzzbar}
\end{figure}

To see why this is the case, we note that $\Delta/w \ll 1$, the
value of $O_{22}^{\rm DE}$ is expected to be identical irrespective
of whether the starting state is $|{\mathbb Z}_2\rangle$ or $|{\bar
{\mathbb Z}_2}\rangle$. This situation is highlighted in the left
panels of Fig.\ \ref{figzzbar}. The top left panel shows the
expectation value $O_{22}^{(m)}= \langle m|\hat n_2 \hat n_4
|m\rangle$ corresponding to the Floquet eigenstate $|m\rangle$ as a
function of their dimensionless quasienergy $\phi=E_F^{(m)}T/
\hbar$. The bottom panel shows the overlap of the eigenstates
$|m\rangle$ with the initial states. For small $\Delta/w=0.2$, we
find that the largest value of $O_{22}^{(m)}$ occurs for eigenstates
$|m\rangle$ which lie close to the middle of the Floquet spectrum
($E_F \simeq 0$). Furthermore, near the middle of the Floquet
spectrum, some of these states have significant overlap with
$|{\mathbb Z}_2\rangle$ while others have large overlap with $|{\bar
{\mathbb Z}_2}\rangle$ as can be seen the left bottom panel of Fig.\
\ref{figzzbar}. From Eq.\ \ref{o22steady}, we find that this ensures
that $O_{22}^{\rm DE}$ would be almost identical irrespective of
whether the dynamics starts from $|{\mathbb Z}_2\rangle$ or $|{\bar
{\mathbb Z}_2}\rangle$ initial states; in fact they have exactly
identical values at $\Delta=0$.

This situation drastically changes at large $\Delta$, where all the
Floquet eigenstates with large values of $O_{22}^{(m)}$ have
quasienergy $E_F>0$ as shown in the top right panel of Fig.\
\ref{figzzbar}. Moreover, these eigenstates only have large overlap
with $|{\mathbb Z}_2\rangle$. Consequently, the value of
$O_{22}^{\rm DE}$ do not change appreciably for dynamics starting
from $|{\mathbb Z}_2\rangle$ as $\Delta$ is increased; however, its
value reduces drastically if the initial state is $|{\bar {\mathbb
Z}_2 }\rangle$. Thus with increasing $\Delta$, $O_{22}^{\rm DE}$
drops rapidly to subthermal values and approaches zero for the
$|{\bar {\mathbb Z}_2 }\rangle$ initial state; this leads to
qualitatively different $\Delta$-dependence of $O_{22}^{\rm DE}$.

In addition, the top and the middle panel plots of Fig.\ \ref{figss}
indicate the presence of dynamical freezing at special values the
drive frequencies $\omega_D \simeq \lambda/(2 n_0\hbar)$ (where $n_0
\in Z$) for which $w_r\simeq 0$ . At these values, the initial
states $|0\rangle$, $|{\mathbb Z}_2\rangle$ and $|{\bar {\mathbb
Z}_2}\rangle$ becomes approximate eigenstates of the Floquet
Hamiltonian leading to freezing. The neighborhood of these freezing
frequencies is marked by strong deviation from the ETH predicted
value of $O_{22}$ as shown in the top and middle panels of Fig.\
\ref{figss}; this deviation increases with increasing $\Delta$. In
contrast, for $|{\bar{\mathbb Z}_2}\rangle$ initial state and for
$\Delta > w$, the freezing extends over a range of frequencies, as
seen in the bottom panel of Fig.\ \ref{figss}.

\subsection{Oscillations around freezing frequencies}
\label{oscsec}

\begin{figure}
\centering{\ing[width=\linewidth]{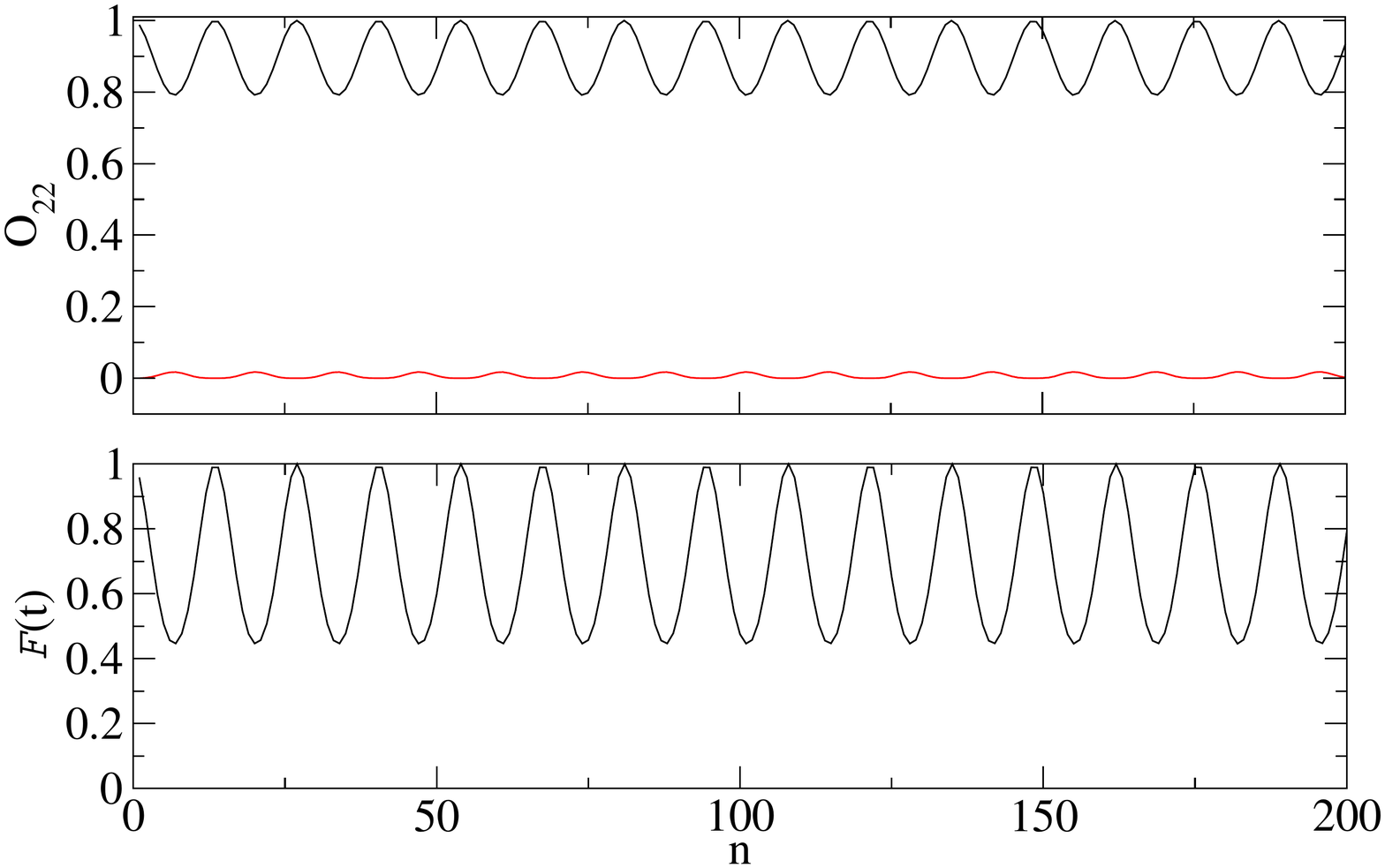}} \caption{Top panel:
Plot of $O_{22}$ as a function of the number of drive cycles $n$
starting from the $|0\rangle$ (red line), $|{\mathbb Z}_2\rangle$
(black line). The bottom panel shows the plot of fidelity ${\mathcal
F}$ as a function of $n$ for $|{\mathbb Z}_2\rangle$ initial state.
For all plots $L=14$, $\lambda/w=15$ and $\hbar \omega_D/w= 6.9$.
All energies(frequencies) are scaled in units of $w$ ($w/\hbar$).
See text for details. } \label{figfr}
\end{figure}
\begin{figure}
\centering{\ing[width=\linewidth]{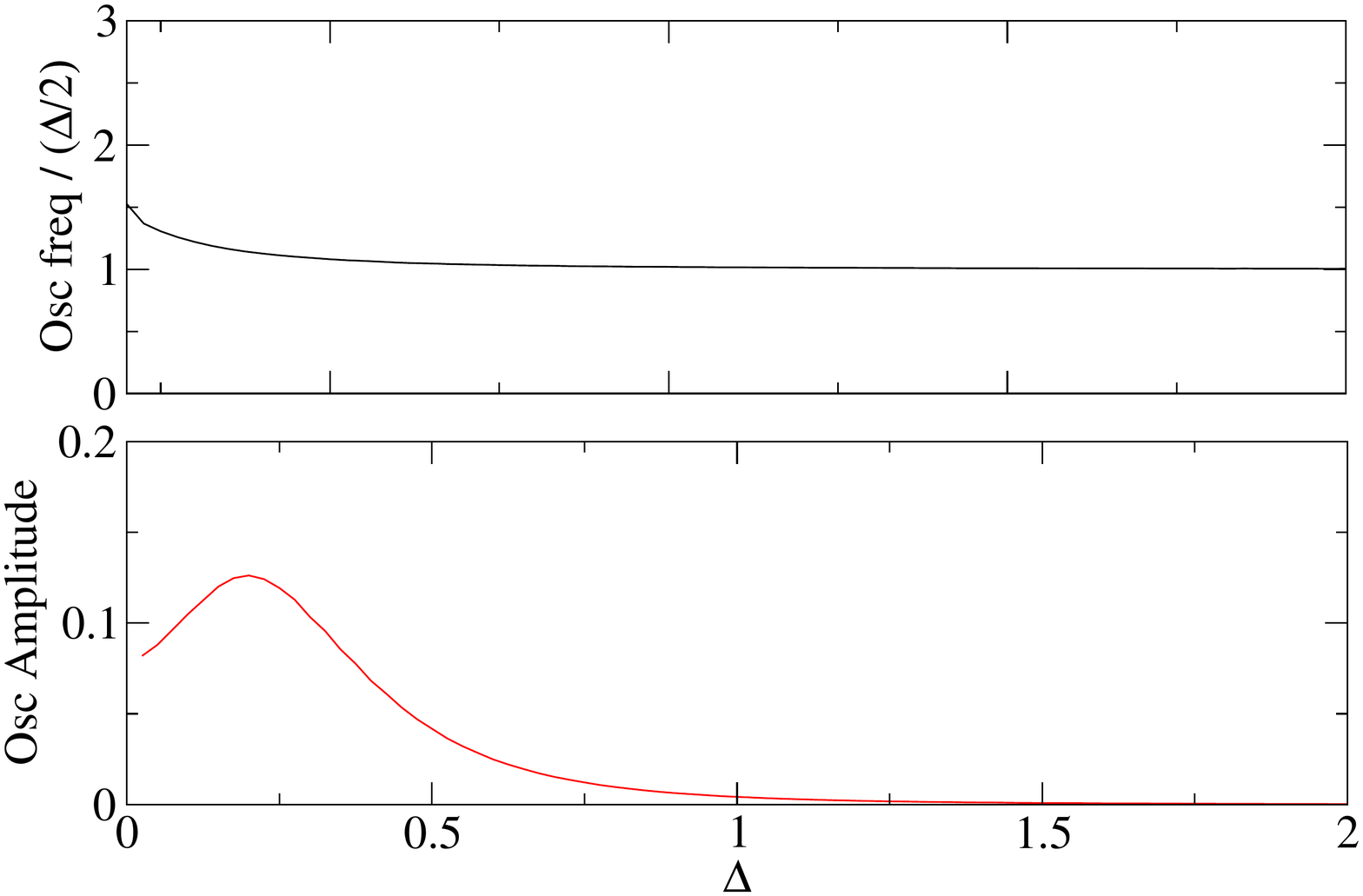}} \caption{Top panel:
Plot of the frequency (top panel) and the amplitude (bottom panel)
of coherent oscillation of $O_{22}$ as a function of $\Delta$ for
$\hbar \omega_D/w=6.9$ and $\lambda/w=15$. All energies(frequencies)
are scaled in units of $w$($w/\hbar$). See text for details. }
\label{figframp}
\end{figure}

The stroboscopic dynamics of $O_{22}$ as a function of $n$ around
the freezing frequencies is shown in the top panel Fig.\
\ref{figfr}. We find that $O_{22}$ does not thermalize; instead, it
displays long time oscillation with a fixed frequency $\omega_{\rm
osc} \simeq \Delta/\hbar$. We have checked this numerically for $n
>1000$ cycles. The amplitude of these oscillation decrease as the
freezing point is approached and also as $\Delta$ is increased; it
is a monotonically decreasing function of $w_r/\Delta$. However, the
period of the oscillations remains constant for $\Delta/w \ge 1$, as
shown in Fig.\ \ref{figframp}. A similar feature is seen from the
fidelity ${\mathcal F} =|\langle {\mathbb Z}_2|\psi(n T)\rangle|^2$,
where $|\psi(nT)\rangle$ denotes the many-body state at time $t=nT$,
as shown in the bottom panel of Fig.\ \ref{figfr}. In contrast, the
oscillation amplitude is close to zero for a wide range of
frequencies around the freezing point when one starts from the
$|{\bar {\mathbb Z}}_2\rangle$ state.

The characteristics of these oscillations at large $\Delta/w_r$ can
be qualitatively understood as follows. Let us consider the
evolution operator when $\Delta_r \gg w_r$. Clearly, in this limit,
one can write
\begin{eqnarray}
U(nT,0) &=& \sum_{\alpha} e^{-i n \epsilon_{\alpha}^F  T/\hbar}
|\alpha\rangle \langle \alpha| \nonumber\\
\epsilon_{\alpha} &=& E_0 + \alpha \Delta + {\rm O}(w_r^2/\Delta^2)
\nonumber\\
|\alpha\rangle &=& \sum_p c_{p;q}^{\alpha} |p;q\rangle
\label{uansatz}
\end{eqnarray}
where $E_0$ denotes the energy of the state $|0\rangle$ for $w_r=0$
and any eigenstate $|\alpha=\alpha_0\rangle$ of the Floquet
Hamiltonian $H_F$ is mapped to a state in the number basis $|p;q
\rangle = |p_0, q_0\rangle$ for $w_r=0$. We note here that the
states $|p_0; q_0\rangle$ have $p_0$ ($q_0$) up-spins distributed on
even (odd) sites of the chain without violation of the constraint;
these are eigenstates of $H_F^{(1)}$ for $w_r=0$ with quasienergy
\begin{eqnarray}
E= E_0+ (q_0-p_0)\Delta.
\end{eqnarray}
Here we shall assume that $c_{p_0+\mu;q_0}^{\alpha_0} \sim {\rm
O}((w_r/\Delta)^{|\mu|})$ for $\mu=\pm 1,\pm 2 ...$; thus for small
$w_r/\Delta$, $|c_{p_0+\mu;q_0}^{\alpha_0}| \ll
c_{p_0;q_0}^{\alpha_0}$ for any $\alpha_0$. A similar relation holds
for $c_{p_0;q_0+\mu}^{\alpha_0}$.

Let us now consider the initial state $|0\rangle$. The wavefunction
after $n$ drive cycles can be obtained as
\begin{eqnarray}
|\psi(nT)\rangle &=& U(n T,0)|0\rangle = \sum_{\alpha}
c_{0;0}^{\alpha \ast} e^{- i \epsilon_{\alpha} n T/\hbar}
|\alpha\rangle
\nonumber\\
&=& \sum_{\alpha} \sum_{p,q}  c_{0;0}^{\alpha \ast} c_{p;q}^{\alpha}
e^{-i \epsilon_{\alpha} p T/\hbar} |p;q\rangle \label{wavT}
\end{eqnarray}
where $c_{0;0}^{\alpha} = \langle p_0=0;q_0=0|\alpha\rangle$. Since
the correlator $O_{22}$ can receive finite contribution for states
with $p_0>2$ up-spins, we find that the leading contribution to
$O_{22}$ is given by
\begin{eqnarray}
O_{22}^{(0)}(nT) \simeq  | c_{0;0}^{0 \ast} c_{2;0}^{2} + c_{0;0}^{1
\ast} c_{2;0}^1 e^{i \Delta n T/\hbar} |^2 \label{022zeroeq}
\end{eqnarray}
where we have retained only the leading order term in $w_r/\Delta$
and used $\epsilon_{\alpha_0} \simeq E_0- (p_0-q_0) \Delta$ (Eq.\
\ref{uansatz}). This indicates that the oscillation amplitudes will
decay with increasing $\Delta$ as  $(w_r/\Delta)^4$; however the
frequency of these oscillation will be pinned to $\Delta$ as long as
$w_r \ll \Delta$. The change in the oscillation frequency is
expected to occur with reduction of $\Delta$ when ${\rm O}((w_r
/\Delta)^2)$ terms in the expression of $\epsilon_{\alpha}$ becomes
significant. The validation of this argument can be found in Fig.\
\ref{figframp}. We note that for $\Delta_r \gg w_r$, the presence of
a large on-site term which is diagonal in the Fock basis leads to
long-time coherent oscillations; in this case, we do not find the
rapid thermalization expected by ETH.

The coherent oscillations discussed above are expected to have
larger amplitude provided we start from the state $|{\mathbb
Z}_2\rangle$. To see this, we note that in this case, the
wavefunction after $n$ drive cycles is given by
\begin{eqnarray}
|\psi'(nT)\rangle &=&  \sum_{\alpha} \sum_{p,q} c_{L/2;0}^{\alpha
\ast} c_{p,q}^{\alpha} e^{-i \epsilon_{\alpha} n T/\hbar}
|p;q\rangle \label{z2wav}
\end{eqnarray}
where $c_{L/2;0}^{\alpha}= \langle {\mathbb Z}_2|\alpha\rangle$. In
this case, the leading order contribution to $O_{22}$ reads
\begin{eqnarray}
O_{22}^{({\mathbb Z}_2)}(nT) \simeq  | |c_{L/2,0}^{L/2 \ast}|^2 +
c_{L/2,0}^{L/2-1 \ast} c_{L/2-1,1}^{L/2-1} e^{i \Delta n T/\hbar}
|^2 \label{z2amposc}
\end{eqnarray}
which leads to ${\rm O}(w_r^2/\Delta^2)$ oscillation amplitude.

In contrast, if one starts from the $|{\bar {\mathbb Z}_2}\rangle$
states, the oscillation amplitude is expected to be vanishingly
small. This is due to the fact that $|{\bar {\mathbb Z}_2}\rangle$
has insignificant overlap with any $|p_0;q_0\rangle$ state which
contributes to $\langle \hat n_2 \hat n_4\rangle$. We also note that
density-density correlation between odd sites (such as $O_{12}$)
would show similar behavior for the initial state $|0\rangle$;
however, it would have vanishing oscillation amplitude for the
$|{\mathbb Z}_2\rangle$ initial state instead of $|{\bar {\mathbb
Z}_2}\rangle$.

\section{Crossover from weak to strong staggered detuning}
\label{crover}

In this section, we shall study the behavior of the quasienergy
eigenstates as $\delta E_2/w_r$ is gradually increased. We note at
the outset that this behavior can be reproduced by studying the
properties of eigenstates of
\begin{eqnarray}
H_m &=& -(\Delta'/2) \sum_j (-1)^j \sigma_j^z - \sum_j \tilde
\sigma_j^x  \label{hmham}
\end{eqnarray}
such that $\Delta' \equiv 2 \delta E_2/w_r$. This is due to the fact
that a variation of $\omega_D$ in $H_F^{(1)}$ can take us close to
both the commensuration point where $\delta E_2 \simeq 0$ (for
example, for $\Delta \simeq 1.2$ and $\hbar \omega_d \simeq 3.6$)
and the dynamical freezing point where $w_r \simeq 0$ (for example,
around $\lambda T/\hbar = 4 \pi$). We therefore expect this analysis
to reveal the nature of the crossover between Hilbert space
structure having quantum scars ($\Delta'=0$) to that having
clustering ($\Delta' \gg 1$). Our numerical results are discussed in
Sec.\ \ref{scarnum} while a FSA based calculation for scars at small
and intermediate value of $\Delta$ is presented in Sec.\
\ref{scarfsa}.

\subsection{Numerical results for scars}
\label{scarnum}

\begin{figure}
\centering{\ing[width=\linewidth]{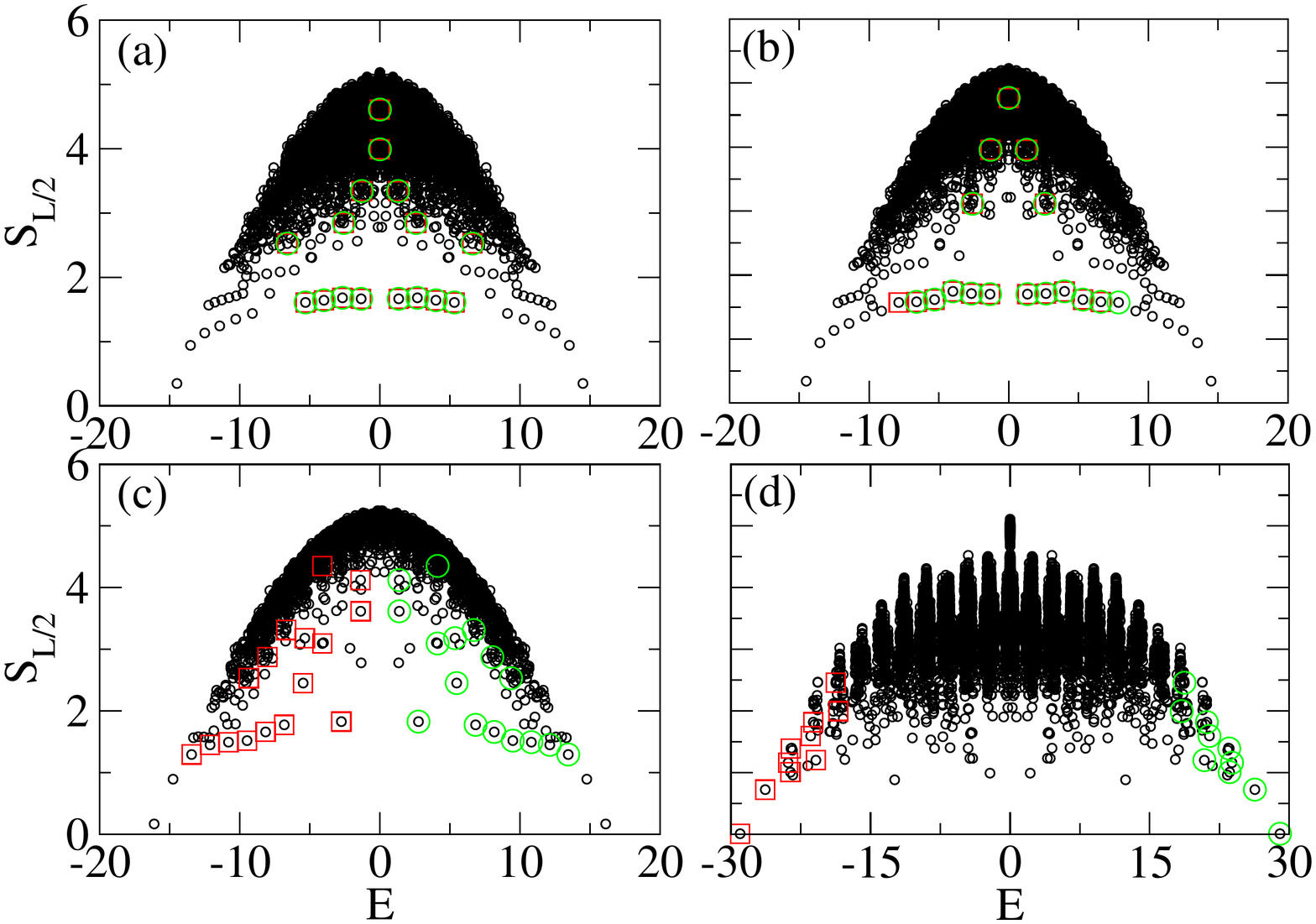}} \caption{Plot of the
half chain entanglement $S_{L/2}$ of the eigenstates of $H_m$ as a
function of energy $E$ for (a) $\Delta'=0.005$, (b) $0.05$, (c)
$0.5$, and (d) $2$. The green and the red circles indicate states
with high overlap with $|{\bar {\mathbb Z}_2}\rangle$ and $|{\mathbb
Z}_2\rangle$ respectively. All plots have same Y axes range and
correspond to $L=24$. See text for details. } \label{figcren1}
\end{figure}

\begin{figure}
\centering{\ing[width=\linewidth]{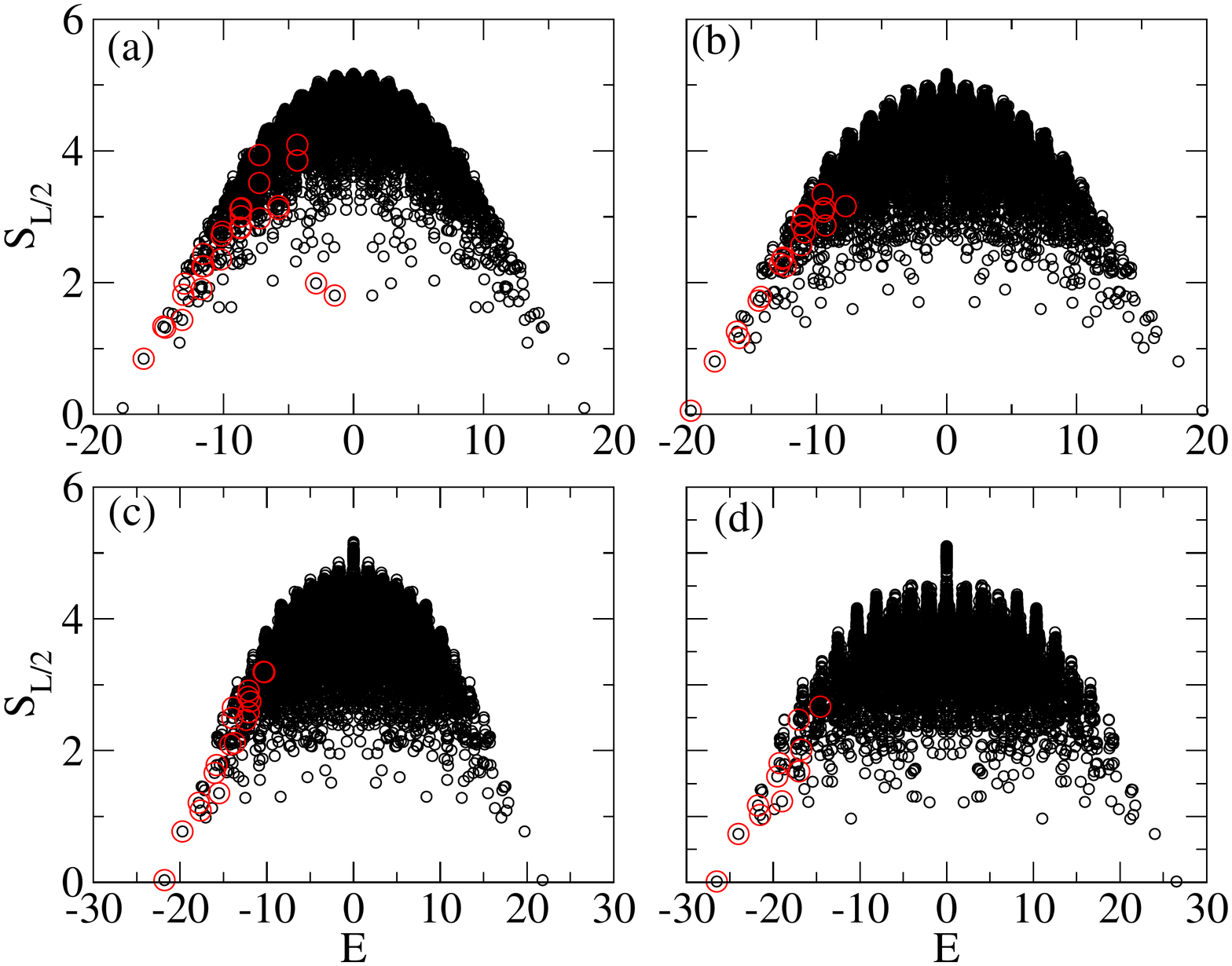}} \caption{Plot of the
half chain entanglement $S_{L/2}$ of the eigenstates of $H_m$ as a
function energy $E$ for (a) $\Delta'=0.75$, (b) $1$, (c) $1.25$, and
(d) $1.75$. The red circles indicate states with high overlap with
$|{\mathbb Z}_2\rangle$. All plots have same Y axes range and
correspond to $L=24$. See text for details. } \label{figcren2}
\end{figure}

In this section, we first show the entanglement entropy of the
eigenstates of $H_m$ as a function of dimensionless energy $E$ for
four representative values of $\Delta'$ in Fig.\ \ref{figcren1}. The
top left panel of Fig.\ \ref{figcren1} shows the existence of
quantum scars for small $\Delta'\ll 1$. These scars which occurs
close to the commensuration point where $x\ll 1$, have, in
accordance with standard expectation, almost equal overlap with
$|{\mathbb Z}_2\rangle$ (red circles) and $|{\bar {\mathbb
Z}_2}\rangle$ (green circles) states. We find as we increases
$\Delta'$, the scars with $E_F <0$ develops higher overlap with
$|{\mathbb Z}_2\rangle$ while those with $E_F>0$ exhibits higher
overlap with $|\bar {\mathbb Z}_2\rangle$. This asymmetry can be
understood from the fact that the presence of a large $\Delta'$
pushes the states with high $|{\mathbb Z}_2\rangle$ ($|{\bar
{\mathbb Z}_2}\rangle$) overlap to lower (higher) quasienergy. For
$\Delta' \gg 1$, there are almost no mid-spectrum states which has
high overlap with either $|{\mathbb Z}_2\rangle$ or $|{\bar {\mathbb
Z}_2}\rangle$. Remarkably, even in this regime, we find existence of
scar like states as seen from bottom right panel of Fig.\
\ref{figcren1}; these do not show high overlap with either $|{\bar
{\mathbb Z}_2}\rangle$ or $|{\mathbb Z}_2\rangle$. The crossover
between these two situation happens between $0.75 \le \Delta'\le 1$
as can be seen from Fig.\ \ref{figcren2}. Thus at these values of
$\Delta'$, an almost continuous band of scar states remain although
their nature evolves with increasing $\Delta'$.

\begin{widetext}
\begin{figure*}
\centering{\ing[width=0.49\linewidth]{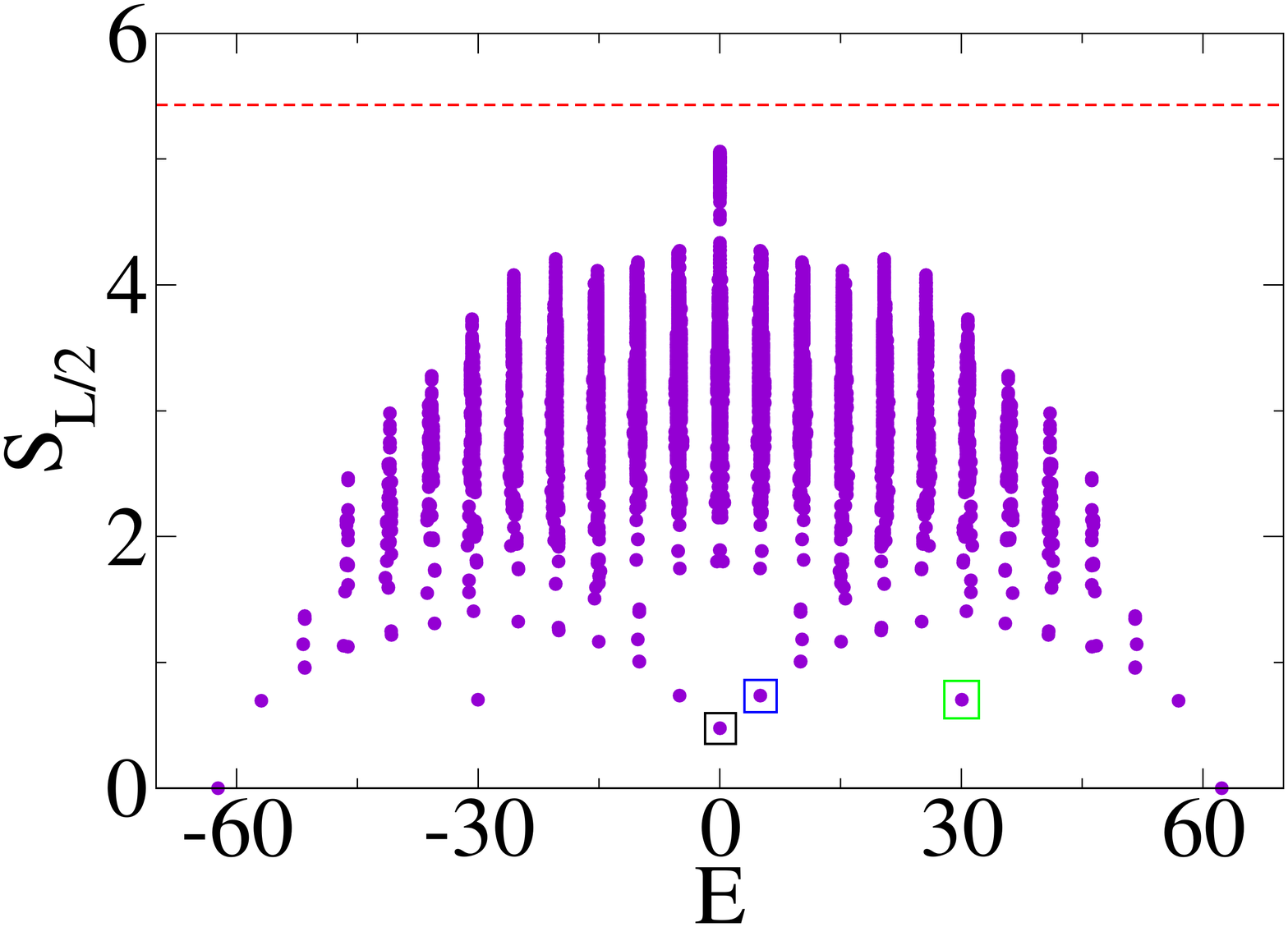}}
\centering{\ing[width=0.49\linewidth]{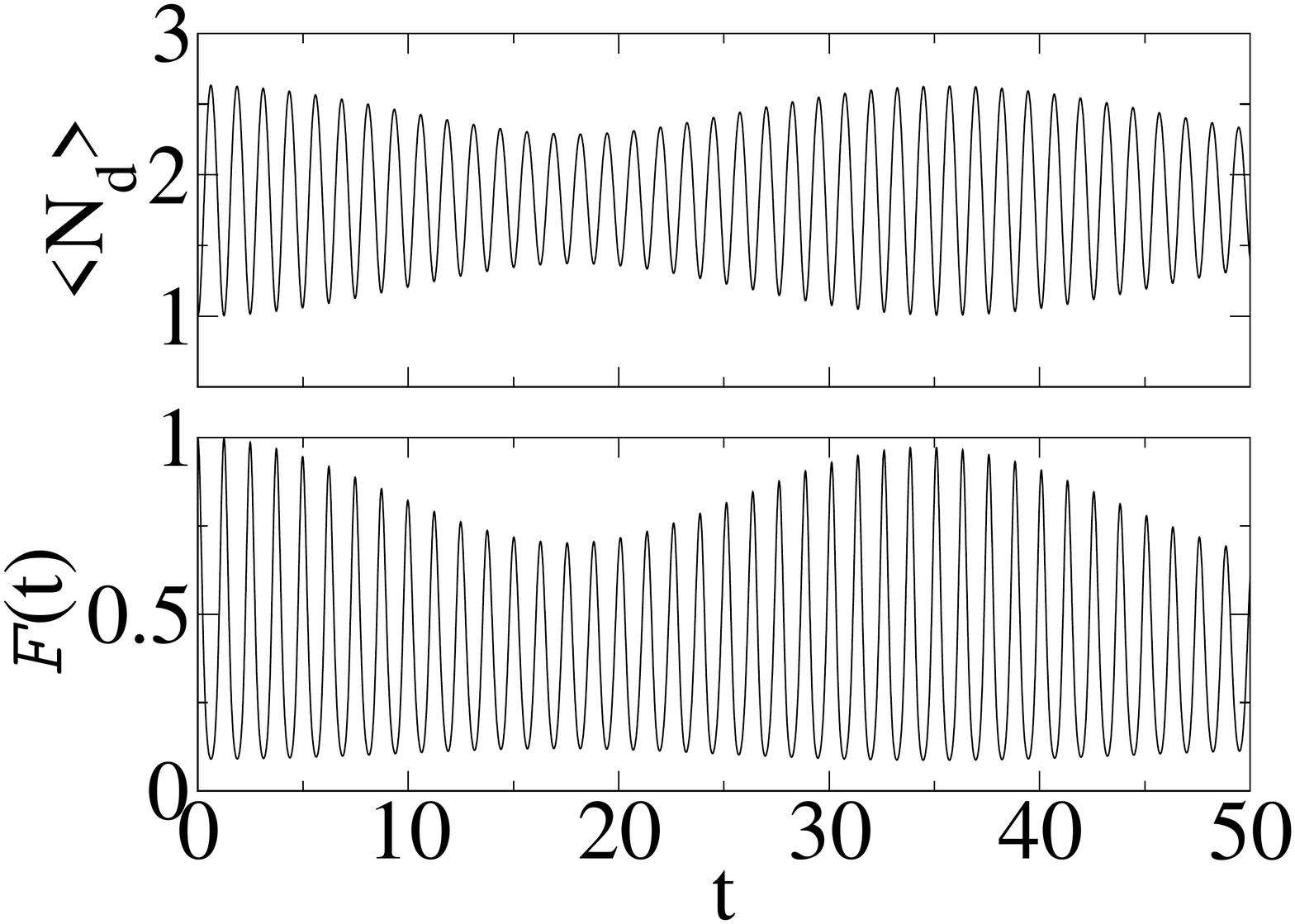}} \caption{Left:
Plot of $S_{L/2}$ for eigenstates of $H_m$ showing the scar states
having large overlap with $|0\rangle$ (black square), $|1\rangle$
(blue square), and $|{\mathbb Z}_4\rangle$ (green square). The red
dashed line indicates the ETH predicted Page value of $S_{L/2}$.
Right Panel: The dynamics of $\langle N_d \rangle $ starting from
state $|1\rangle$ (top) and the fidelity $F(t)=|\langle
\psi(t)|1\rangle|^2$ (bottom) as a function of time showing scar
induced long-time coherent oscillations. For all plots $L=24$ and
$\Delta'=5$. See text for details.} \label{figschi}
\end{figure*}
\end{widetext}

This structure of scars is preserved at higher values of $\Delta'$
where the Hilbert space is strongly clustered. The mid-spectrum
states here do not have significant overlap with either $|{\mathbb
Z}_2\rangle$ and $|{\bar {\mathbb Z}_2}\rangle$. Instead, as we find
numerically and as shown in the left panel of Fig.\ \ref{figschi},
two of these states have high overlap with $|{\mathbb Z}_4\rangle =
|\uparrow, \downarrow,\downarrow,\downarrow, \uparrow,
\downarrow,\downarrow,\downarrow ...\rangle$ states. Other mid
spectrum scar states have strong overlap either with the single
Rydberg excited state $|1\rangle = |\uparrow, \downarrow, \downarrow
....\rangle$ or the vacuum state $|0\rangle$ as shown in the left
panel of Fig.\ \ref{figschi}.

\begin{figure}
\centering{\ing[width=\linewidth]{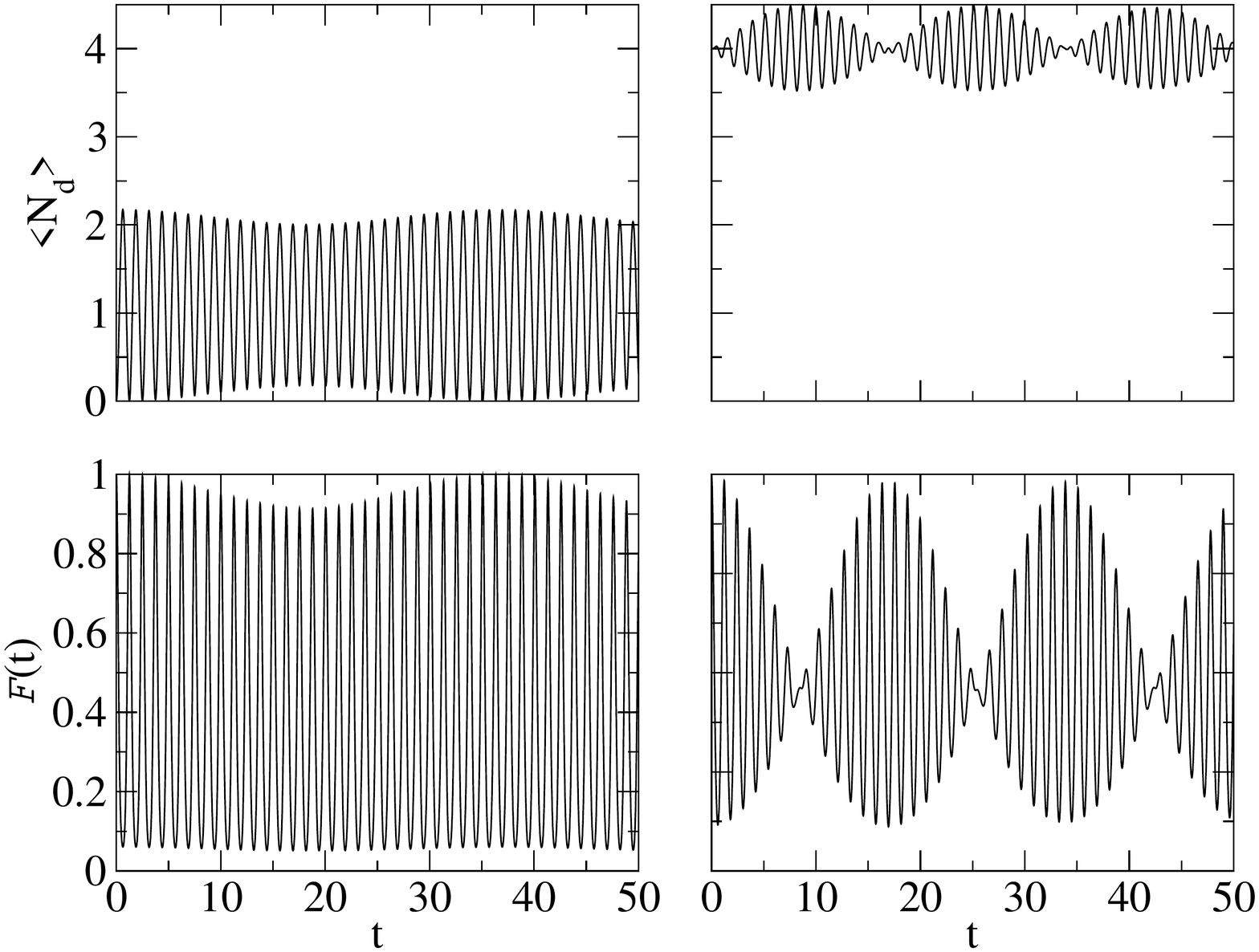}} \caption{Left: The
dynamics of $\langle N_d \rangle $ (top) starting from the initial
state $|0\rangle$ and the fidelity $F(t)=|\langle
\psi(t)|0\rangle|^2$ (bottom) as a function of time. Right: Similar
plots for the initial $|{\mathbb Z}_4\rangle$ state. All left panel
plots have same Y axes range as their right panel counterparts. All
plots correspond to $L=16$ and $\Delta'=5$. See text for details.}
\label{figscdyn}
\end{figure}
These mid-spectrum scar states have no analogue in their
counterparts for small $\Delta'$. Indeed this can be seen by
studying the dynamics of $\langle N_d \rangle  =\langle \sum_j
(\sigma_j^z+1)/2\rangle$ as a function of time. The fidelity $F_1=
|\langle 1|\psi(t)\rangle|^2$ of these oscillations indicate perfect
revivals as can be seen from the bottom right panel of Fig.\
\ref{figschi}. The presence of such oscillations starting from
single spin-up state could not have been due to the $|{\mathbb
Z}_2\rangle$ scars. Similar oscillatory coherent dynamics with near
perfect fidelity revivals is also seen in this regime for initial
states $|0\rangle$ and $|{\mathbb Z}_4\rangle$ as shown in Fig.\
\ref{figscdyn}.

\begin{figure}
\centering{\ing[width=\linewidth]{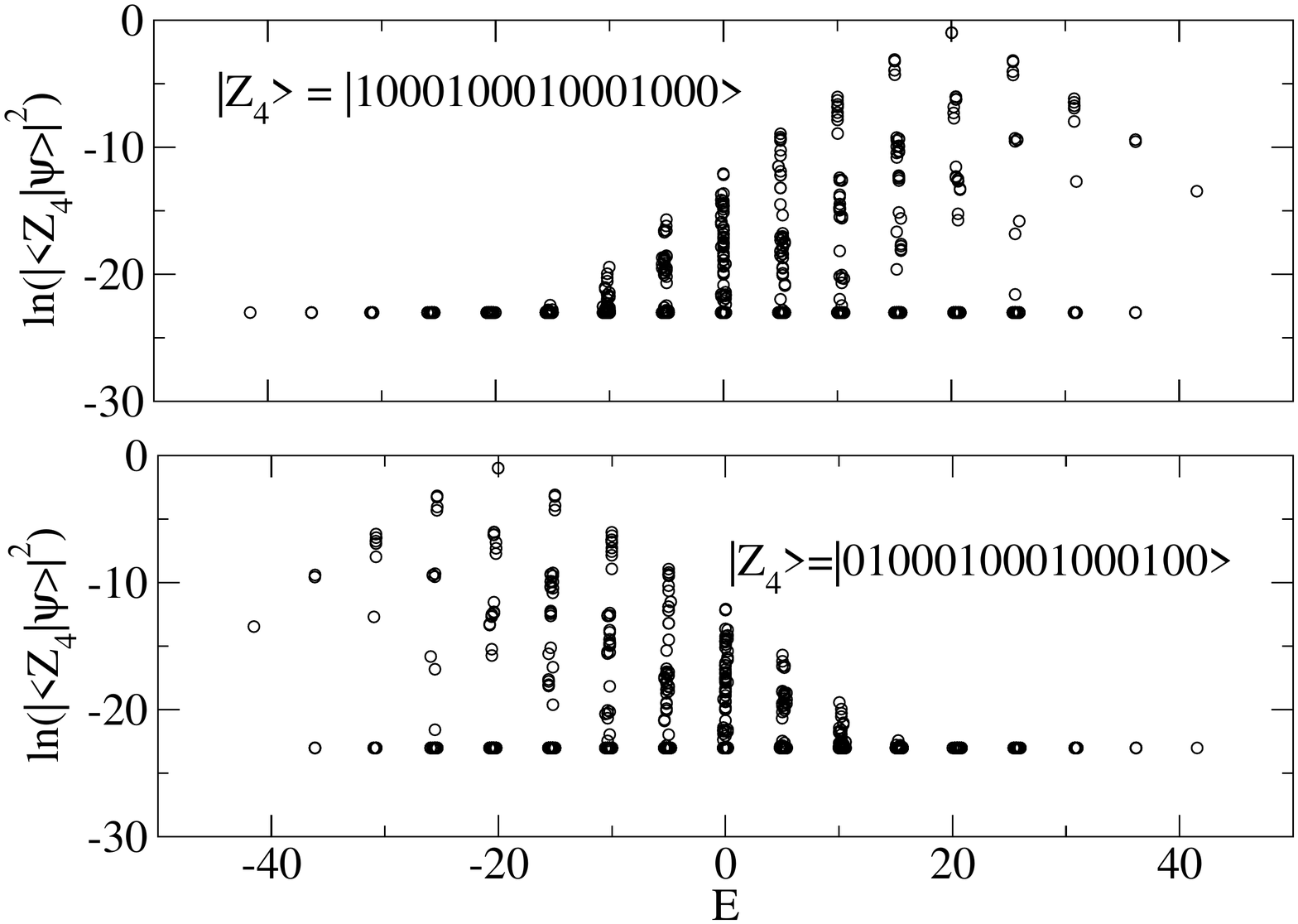}} \caption{Plot of $\ln
|\langle {\mathbb Z}_4|n\rangle|^2$ as a function of $E_F$ for
$|{\mathbb Z}_4\rangle$ states which hosts Rydberg excitations on
even (top panel) and odd (bottom panel) sites. For all plots
$\Delta'=5$ and $L=16$. See text for details.} \label{figoverlap}
\end{figure}

The frequencies of these oscillations and the nature of the dynamics
starting from the $|{\mathbb Z}_4\rangle$ state can be understood as
follows. From the plot of the overlap of the initial state with the
Floquet eigenstates, shown in Fig.\ \ref{figoverlap}, we find that
the overlap of $|{\mathbb Z}_4\rangle$ is maximal with the one of
the mid-spectrum scar states. This state has $E_F>0$ ($E_F<0$) if
the $|{\mathbb Z}_4\rangle$ hosts Rydberg excitations on odd(even)
sites; this is due to the presence of a large $\Delta'$ in the
Hamiltonian which breaks the particle-hole symmetry. The next
largest overlap corresponds to two states on both sides of the
$|{\mathbb Z}_4\rangle$ states; these have slightly different
energies and belong to primary clusters characterized by $Z_{\pi}=5$
and $7$ respectively. The difference in energy between the $Z_4$
scar and these two states leads to the main oscillation frequency
$\omega \simeq \Delta'/(2 \hbar)$; in addition, the small difference
in frequencies of these two states lead to a long-time beating
phenomenon as can be seen from the right panel Fig.\ \ref{figscdyn}.
The analysis of the dynamics starting from $|1\rangle$ or
$|0\rangle$ reveals a qualitatively similar picture and we do not
discuss them in details here.

The nature of the scar states with large overlap with $|1\rangle$
states may be qualitatively understood by noting the form of
$H_4^{(1)}$ given in Eq.\ \ref{pertapp4}. For strong $\Delta'$,
where the analysis of Ref.\ \onlinecite{ethv05b} holds, $H_4^{(1)}$
is the leading term in $1/\Delta'$ which induces spin dynamics. Here
we note that such a term acting on a state $|1_j\rangle$ (which
represents a single Rydberg excitation on the $j^{\rm th}$ site)
connects it to $|1_{j+2}\rangle$. Thus it creates an almost closed
subspace of $L/2$ such states; the corresponding Floquet eigenstates
is therefore expected to have low entanglement and could be a
candidate for the scar state having large overlap with $|1\rangle$.

A similar, but more complicated holds for scars having large overlap
with $|{\mathbb Z}_4\rangle$. We note that $H^{(1)}_4$ acting on
$|{\mathbb Z}_4\rangle$ connects it to a series of $L/4$ states
where any up-spin state is shifted by two sites. These states are
denoted by $|{\mathbb Z}_{4j}\rangle$ and has repeated block of one
up-spin on site $4\ell$ for integer $\ell$ and three subsequent
down-spins. Now let us consider the subsequent action of $H_4^{(1)}$
on this state; it is easy to see that a further shift of the $j^{\rm
th}$ up-spin is either forbidden by the constraint or gives back
$|{\mathbb Z}_4\rangle$. This ensures that action of $H^{(1)}_4$ on
$|{\mathbb Z}_{4j}\rangle$ leads to $^{L/4}C_2$ new states.
Following this argument we find that the dimension of the almost
closed subspace formed by the action of $H_4^{(1)}$ on $|{\mathbb
Z}_4\rangle$ is $\sim \sum_{n=1 ..L/4} \,^{L/4}C_n$. For finite $L
\le 24$ studied in our numerics, this leads to a small, almost
closed, subspace of states. Thus qualitatively one may expect to
have a low-entangled eigenstate formed out of their superposition
which have significant overlap with $|{\mathbb Z}_4\rangle$. We
conjecture that such a state may be the $Z_4$ scar that we find. We
also note that this suggests that scars having large overlap with
$|1\rangle$ (for which the closed subspace always has ${\rm O}(L)$
states) would be more stable for large $L$ than their counterparts
having large overlap with $|{\mathbb Z}_4\rangle$.

\subsection{FSA calculations}
\label{scarfsa}

To understand the nature of these scars at low and intermediate
$\Delta'$ and the change in their properties as $\Delta'$ is
increased, we use a semi-analytic FSA treatment. To this end, we
first note that the presence of the staggered detuning term suggests
that such a construction should distinguish between odd and even
sites in the lattice. Therefore we adapt the formalism of Ref.\
\onlinecite{turpap1} which allows for such a sublattice resolved
formulation. In this method, the states are labeled by a set of two
integers $n_1$ and $n_2$ which denotes number of Rydberg excitations
(up-spins) on even and odd sites respectively. In what follows we
shall construct the matrix elements of $H_m$ between two such states
and then diagonalize this matrix numerically. Such a class of states
constitutes a symmetric subspace defined from a set of equivalence
classes $(n_1,n_2)$; all elements of these classes are invariant
under shuffling of excitations in each sublattice \cite{turpap1}.
Thus in this scheme, the action of translation by an even number of
lattice states leads to an equivalent state. The assumption here is
that the key information about properties of the states reside in
the number of excitation in each sublattice. We note that this
procedure naturally yield states which have significant overlap with
$|{\mathbb Z}_2\rangle$ and $|{\bar {\mathbb Z}_2}\rangle$ since
these Neel states are included as elements in the subspace
\cite{turpap1}.

\begin{figure}
\centering{\ing[width=\linewidth]{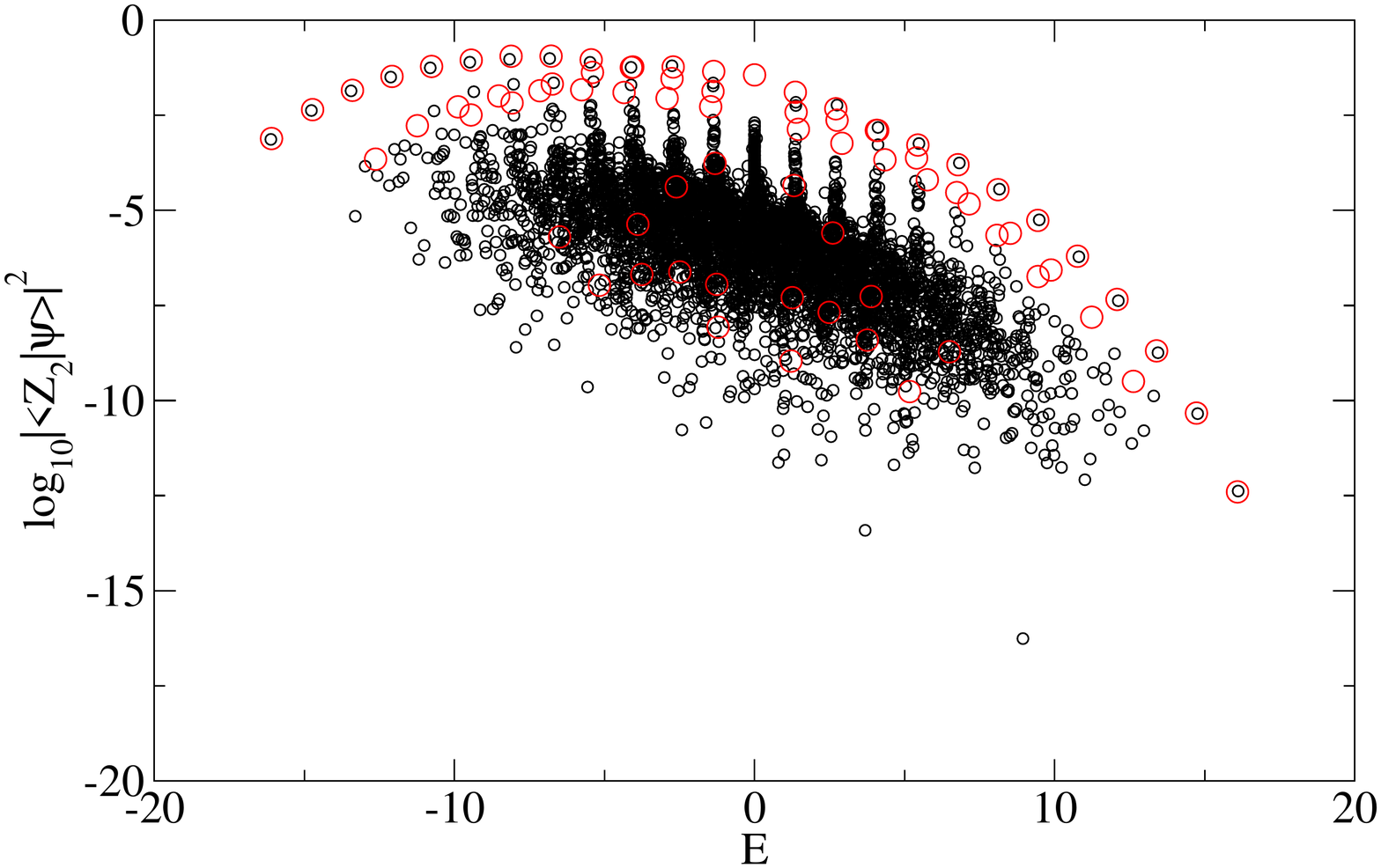}} \caption{A plot of
$\log_{10} |\langle{\mathbb Z}_2|\psi\rangle|^2$ for states obtained
via FSA (large red circles) and those obtained from exact numerics
(small black circle) for $\Delta'=0.5$ and $L=24$. See text for
details.} \label{figfsa1}
\end{figure}
The first step towards construction of the matrix elements of $H_m$
is to construct an orthonormal basis for the symmetric subspace. To
this end, we use the prescription of Ref.\ \onlinecite{turpap1} and
write, for a chain with $2L$ sites
\begin{eqnarray}
|n_1, n_2\rangle &=& \frac{1}{\mathcal N} \sum_a  |\alpha_a\rangle
\label{fsabasis} \\
{\mathcal N} &=& \frac{(L-n_1-n_2)L}{(L-n_1)(L-n_2)} \left(
\begin{array} {c} L-n_1 \\ n_2 \end{array} \right) \left(
\begin{array} {c} L-n_2 \\ n_1 \end{array} \right) \nonumber
\end{eqnarray}
where any state $|\alpha_a\rangle$ contain $n_1$ and $n_2$ total
Rydberg excitations on even and odd sites respectively. The number
of such states, ${\mathcal N}$, depends on specific values of $n_1$
and $n_2$ and also the Hilbert space constraint of having no two
Rydberg excitation on neighboring sites.

The matrix element of $H_m$ between these states can be easily
obtained following the prescription of Ref.\ \onlinecite{turpap1}.
To this end, we write, using $|0\rangle$ as the reference state,
\begin{eqnarray}
&& \langle n_1, n_2| H_m |m_1,m_2\rangle = F_1+F_2 \nonumber\\
&&F_1 = \Delta' (n_2-n_1) \delta_{n_1, m_1} \delta_{n_2, m_2}
\label{fsamatel} \\
&&F_2 = - \Big[M(n_1,n_2) \delta_{m_1,n_1+1} \delta_{m_2, n_2}
\nonumber\\
&& \quad \quad + M(n_2,n_1) \delta_{m_1 n_1} \delta_{m_2,n_2+1}
\Big] +{\rm h.c.}
\nonumber\\
&&M(n_1,n_2) = \sqrt{\frac{(L-n_1 -n_2)(L-n_1-n_2+1)n_1}{L-n_1}}
\nonumber
\end{eqnarray}
We note that the structure of the FSA Hamiltonian suggests that a
large $\Delta'$ essentially separates out states with large $n_1$
from those with large $n_2$. Thus the eigenvalues of these
Hamiltonian, for large $\Delta'$, is expected to have large overlap
with either $|{\mathbb Z}_2\rangle$ or $|{\bar{\mathbb Z}_2}\rangle$
but not both.

\begin{figure}
\centering{\ing[width=\linewidth]{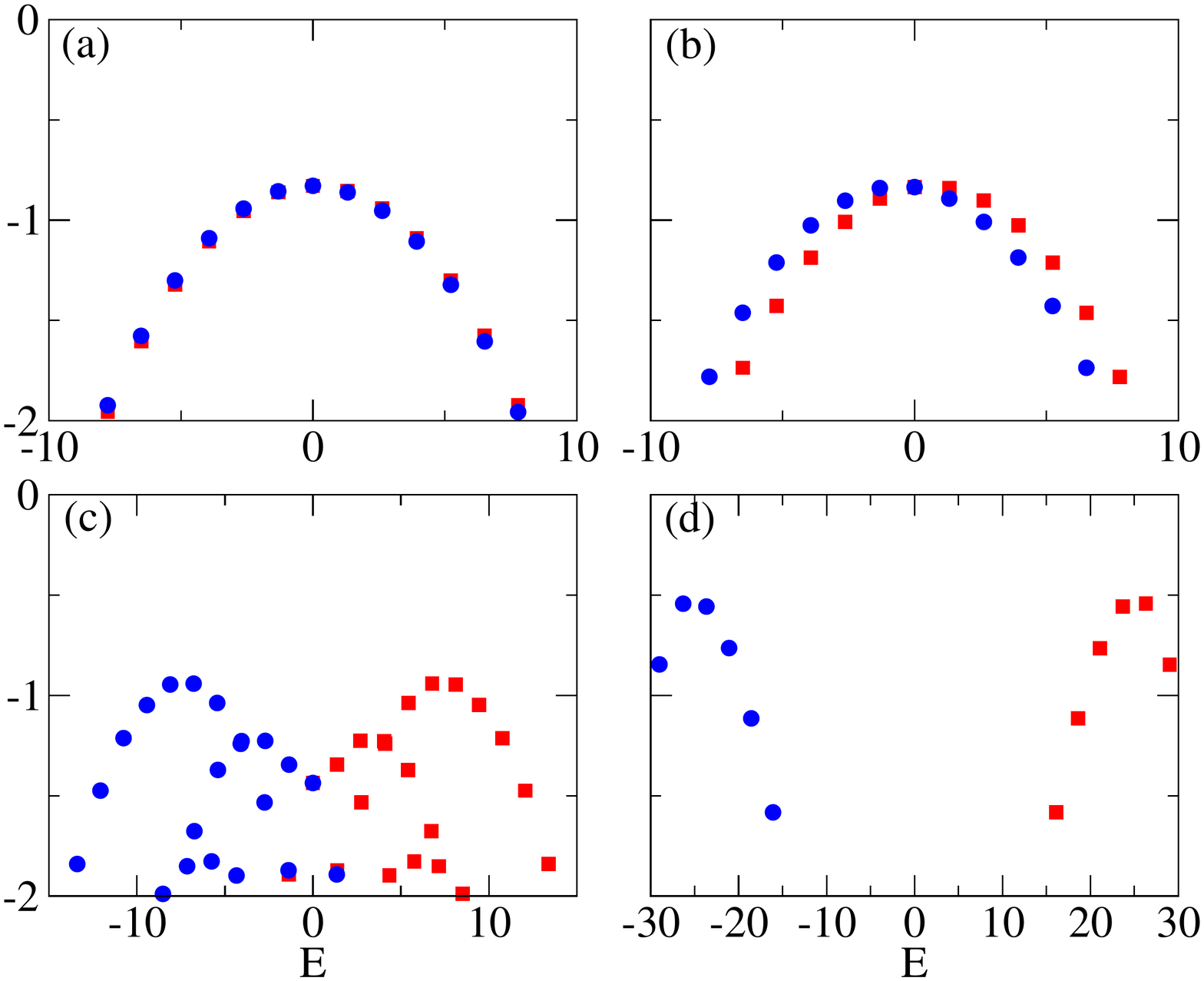}} \caption{A plot of
$\log_{10} |\langle{\mathbb Z}_2|\psi\rangle|^2$ (blue circles) and
$\log_{10} |\langle {\bar {\mathbb Z}_2}|\psi\rangle|^2$ (red
circles) for states obtained via FSA for (a) $\Delta'=0.005$, (b)
$0.05$, (c) $0.5$, and (d) $2$. All plots have same Y axes range and
correspond to $L=24$. See text for details.} \label{figfsa}
\end{figure}
The diagonalization of the FSA matrix constructed from Eq.\
\ref{fsamatel} leads to the eigenvalues of $H_m$ for a fixed
$\Delta'$. A comparison of the $|{\mathbb Z}_2\rangle$ overlap of
these states with those of their counterparts obtained via exact
numerics for $L=24$ and $\Delta'=0.5$, plotted in Fig.\
\ref{figfsa1}, shows a near-exact match between the scar eigenstates
obtained by these two methods; the eigenstates with $E_F<0$ has a
much higher overlap with $|{\mathbb Z}_2\rangle$ compared to those
with $E_F>0$ in both cases. This allows us to surmise that the
present FSA captures the essence of the scar eigenstates in the
presence of finite $\Delta$.

The overlap between these eigenstates with $|{\mathbb Z}_2\rangle$
(red dots) and $|{\bar {\mathbb Z}_2}\rangle$ (blue dots) is shown
in Fig.\ \ref{figfsa}. From these, we find that the FSA eigenstates
qualitatively reproduce the characteristics of scars seen in exact
numerics. For low $\Delta'$, the scar states have near exact overlap
with both the Neel states (top left panel of Fig.\ \ref{figfsa}); in
contrast, as $\Delta'$ is increased, states with $E_F>0$ develop a
large overlap with $|{\bar {\mathbb Z}_2}\rangle$ while those with
$E_F <0$ overlaps strongly with $|{\mathbb Z}_2\rangle$. Thus the
crossover of the nature of the scars as a function of $\Delta'$ is
well captured by the FSA. This behavior can be understood to be a
consequence of the presence of the diagonal elements of the FSA
matrix for finite $\Delta'$ (Eq.\ \ref{fsamatel}); their presence
pushes states having large overlap with $|{\mathbb Z}_2\rangle$
($|{\bar {\mathbb Z}_2}\rangle$) to opposite ends of the Floquet
spectrum. The mid-spectrum scar states at large $\Delta'$, however,
can not be captured by the present FSA formalism. A more detailed
quantitative understanding of such states and their description in
terms of a modified FSA approach is left for future studies.

\section{Discussion}
\label{diss}

In this work, we have studied a periodically driven Rydberg chain
with both uniform and staggered detuning. We have primarily focussed
on the limit of large drive amplitude and a square pulse protocol;
the uniform detuning term has been made time dependent. We find that
the presence of a staggered detuning term leads to several features
of the driven system that have no analogue in Floquet dynamics of
Rydberg atoms without such a term.

The first of such features involves ergodicity violation of such
system via clustering of Floquet eigenstates. We find that when
$\Delta \gg w_r$, the eigenstates of the Floquet Hamiltonian arrange
themselves into several discrete well-separated clusters in the
Hilbert space. Each of these clusters belong to definite values of
$Z_{\pi}$ and the number of such clusters depend on the ratio of
$\hbar \omega_D/\Delta$. We dub this phenomenon as primary
clustering and recognize this to be consequence of
near-integrability of $H_F$ for large $\Delta/w_r$. In addition, we
also find a secondary clustering which leads to discrete set of
states within each of these primary clusters. We tie the presence of
such secondary clustering to the existence of emergent conserved
quantity, $Y$, which was noted earlier in the context of non-driven
Rydberg chains with such staggered detuning in Refs.\
\onlinecite{ethv05a,ethv05b}. We note that these clusterings leading
to violation of ETH do not stem from standard Hilbert space
fragmentation \cite{hsf1,hsf2,hsf3,hsf4} since the Floquet
Hamiltonian of the staggered Rydberg chain do not have dynamically
disconnected sector at any finite $w_r$. Furthermore such clustering
is not expected to survive in the thermodynamic limit; however, they
will be a feature of finite-sized chains which are typically
realized in standard experiments \cite{exp3,exp4}.

We also show that the existence of $Y$ and hence secondary
clustering is contingent on incommensuration of the drive frequency
and $\Delta$ which leads to a non-zero $\delta E_2$. Thus at
commensurate drive frequencies where $\delta E_2 =0$, the secondary
clustering is destroyed and states within each primary cluster
become ergodic. This is reflected in the values of local observables
such as $O_{22}$ near these commensurate drive frequencies which
reach very close to their ETH predicted values. This leads to a
possibility of tuning ergodicity properties of the driven system
using the drive frequency.

We also study the behavior of the correlation function $O_{22}$ for
the driven chain starting from the vacuum and the Neel states. We
find, for initial $|0\rangle$ and $|{\mathbb Z}_2\rangle$ dynamical
freezing of $O_{22}$ at specific drive frequencies whose values are
well approximated by the analytic Floquet Hamiltonian obtained using
FPT. When driven close to these freezing frequencies, $O_{22}$
starting from $|0\rangle$ or $|{\mathbb Z}_2\rangle$ displays
oscillations with perfect revivals. The amplitude of these
oscillations decreases with increasing $\Delta$ in the large or
intermediate $\Delta/w_r$ regime; in contrast, their frequencies are
pinned to $\Delta/\hbar$ for both the initial states. We provide a
simple qualitative explanation of these features using the structure
of the Floquet eigenstates in this regime. In contrast such
oscillations are completely suppressed when the dynamics starts from
$|{\bar {\mathbb Z}_2}\rangle$; this leads to a range of drive
frequencies where one encounters dynamic freezing. This range
increases with $\Delta$. We explain this phenomenon qualitatively
and point out that the dynamics of $O_{j2}$, where $j$ is odd, would
show exactly opposite behavior; it would show range of freezing
frequencies for dynamics starting from $|{\mathbb Z}_2\rangle$
state. This dichotomy is a direct consequence of breaking of the
sublattice symmetry by the staggered detuning term.

We have also studied the quantum scars in the Floquet eigenspectrum
of such a driven Rydberg chain. Such scars are known to exist for
$\Delta'=0$ \cite{ethv1} and are dubbed as PXP scars \cite{ethv01};
they have identical overlap with both the Neel states in this limit.
We find that with increasing $\Delta'$, the scars with energy $E_F
<0$ develop a stronger overlap with $|{\mathbb Z}_2\rangle$ while
those with $E_F>0$ overlap more with $|{\bar {\mathbb Z}_2}\rangle$.
This phenomenon clearly originates from the breaking of the $Z_2$
sublattice symmetry due to presence of $\Delta$. In fact, at large
$\Delta'$, we find that no near mid-spectrum scar states have
significant overlap with either of the Neel states. Instead, as
confirmed from quench dynamics studies in Sec.\ \ref{crover}, they
have a high overlap with either $|{\mathbb Z}_4\rangle$, vacuum
($|0\rangle$) or single dipole ($|1\rangle$) states. This indicates
that the presence of large $\Delta'$ leads to scars which have no
analog with the standard PXP scars studied earlier in the
literature. Our FSA analysis which explains the property of the
scars at weak or intermediate $\Delta'$ fails to capture these
mid-spectrum scars at large $\Delta'$.

In conclusion, we have shown that a driven Rydberg chain with
staggered detuning term leads to several interesting phenomena.
These include ergodicity violation via Floquet eigenstate clustering
in the strong staggered detuning limit, dynamical freezing,
sustained coherent oscillations with perfect revivals near the
freezing frequency, existence of separate class of quantum scars
with large overlap with $|0\rangle$, $|1\rangle$ and $|{\mathbb
Z}_4\rangle$ states, and the possibility of tuning ergodicity
property of these chains with the drive frequency. The experimental
implementation of a Rydberg chain has already been achieved
\cite{exp3,exp4}; a possible extension of some of these experiments
with implementation of staggered detuning may provide a suitable
experimental platform testing our theoretical predictions.

\appendix*

\section{Projection operator formalism}

In this appendix, we detail the computation of $Y$ and $H_2$ using a
perturbative formalism developed in Ref.\ \onlinecite{ethv05b}. To
this end we shall start from a Hamiltonian $H_m$ (Eq.\ \ref{hmham})
in the main text and treat $1/\Delta'$ as the perturbative
parameter. In what follows, we write $H_m= H_0 + H_1$ in terms of
$Z_{\pi}$ (Eq.\ \ref{zeq}). This yields
\begin{eqnarray}
H_0 &=& -\Delta' Z_{\pi}, \quad H_1= -\sum_j \tilde \sigma_j^x.
\label{hamdefpert}
\end{eqnarray}
In the rest of this appendix, we shall treat $H_1$ perturbatively in
the regime where $\Delta'>1$.

The formalism used in Ref.\ \onlinecite{ethv05b} constitutes
construction of a perturbative effective Hamiltonian for such a
system. The method relies on the presence of a canonical
transformation, implemented through an operator $S$, which yields
\begin{eqnarray}
H_{\rm eff} &=& e^{i S} H_m e^{-iS} = H_0 + H^{(1)} + H^{(2)} + ...
\label{pertapp1} \\
H^{(1)} &=& [iS, H_0]+ H_1 \nonumber\\
H^{(2)} &=& P' [iS,H_1] +\frac{1}{2} [iS,[iS,H_0]] P'\nonumber
\end{eqnarray}
where the ellipsis represents higher order terms and $P'$ is a
projection operator which projects the Hamiltonian to the low-energy
manifold of $H_0$ \cite{ethv05b}. The next step is to determine $S$
to first order in perturbation theory by setting $H^{(1)} =0$. A
straightforward calculation using this condition yields
\begin{eqnarray}
iS &=& \frac{1}{2 i \Delta'} \sum_j \tilde \sigma_j^y (-1)^j
\label{pertapp2}
\end{eqnarray}
Using Eq.\ \ref{pertapp2}, one can now compute $H^{(2)}$. While
doing this, it is important to remember that the final results needs
to be projected in the low-energy manifold of $H_0$. This
consideration leads to a single term in the second order which can
be easily computed by substituting Eq.\ \ref{pertapp2} in Eq.\
\ref{pertapp1} given by
\begin{eqnarray}
H^{(2)} &=& \frac{-1}{\Delta'} Y \label{pertapp3}
\end{eqnarray}
where the expression of $Y$ is given in Eq.\ \ref{yexp} of the main
text. We note that as long as the perturbation theory is valid $Y$
emerges as a constant of motion; however, this breaks down, along
with the perturbative approach, when $\Delta'\to 0$.

The higher order terms in the perturbation theory can be computed
systematically using this approach and is detailed out in Ref.\
\onlinecite{ethv05b}. This yields $H^{(3)}=0$; the leading order
non-zero term is given by $H^{(4)}$. As shown in Ref.\
\onlinecite{ethv05b}, this can be written as
\begin{eqnarray}
H^{(4)} &=& H_1^{(4)} + H_2^{(4)}
\nonumber\\
H_1^{(4)} &=& \frac{-1}{2\Delta'^3} \sum_j (-1)^j (\tilde \sigma_j^+
\tilde \sigma_{j+2}^- +{\rm h.c.}) \label{pertapp4} \\
H_2^{(4)} &=& \frac{-1}{2\Delta'^3} \sum_j P_{j-1} \sigma_j^z
P_{j+1} P_{j+2} + P_{j-2} P_{j-}
\sigma_j^z P_{j+1} \nonumber\\
&& - 2P_{j-1} \sigma_j^z P_{j+1} \nonumber
\end{eqnarray}
We note that $H_1^{(4)}$ is the leading order term introducing
non-trivial spin dynamics. The expression of $H_1^{(4)}$ has been
used in Sec.\ \ref{crover} of the main text for developing a
qualitative understanding of the $|{\mathbb Z}_4\rangle$ and
single-dipole scar states.

\end{document}